\documentclass[aps,floatfix,onecolumn,a4paper,noshowpacs, nofootinbib,superscriptaddress,11pt]{revtex4}

\usepackage[utf8]{inputenc}
\usepackage{amsmath,amssymb,graphicx,bm}
\usepackage{geometry}
\makeatletter
\def\@makecaption#1#2{%
  \par\addvspace\abovecaptionskip
  \begingroup
  \small
  \noindent\textbf{#1.} #2\par
  \endgroup
  \addvspace\belowcaptionskip}
\makeatother
\usepackage{siunitx}
\usepackage{hyperref}
\hypersetup{colorlinks=true, linkcolor=blue, citecolor=blue, urlcolor=blue}
\usepackage{graphicx,float,tikz}

\usepackage[all]{xy}
\usepackage{bm,amsmath,upgreek,bigints}

\usepackage{amssymb}
\usepackage{empheq}
\usepackage{color}
\usepackage{epsfig,bm}		
\usepackage{graphicx,epstopdf}
\usepackage{subfigure,upgreek}
\usepackage{pdfpages}
\usepackage{multirow}

\newcommand{\beq}{\begin{eqnarray}}
\newcommand{\eeq}{\end{eqnarray}}

\usepackage[justification=raggedright,singlelinecheck=false]{caption}
\DeclareMathOperator\erf{erf}
\DeclareMathOperator\F{F}
\DeclareMathOperator\erfc{erfc}

\newcommand{\blt}{\textcolor{black}}
\usepackage{bookmark,textgreek}
\usepackage{hyperref,color,xcolor}
\definecolor{lightgray}{gray}{0.95}
\hypersetup{hidelinks,colorlinks=true,breaklinks=true,urlcolor= blue}
\hypersetup{%
    colorlinks = true,
    linkcolor  = blue,
    citecolor = cyan,
  }

\usepackage[dvipsnames]{xcolor}
\usepackage{siunitx}
\definecolor{headergray}{gray}{0.90}

\usepackage{natbib}
\setcitestyle{square,numbers}

\newcommand\orcidroldao{{\href{https://orcid.org/0000-0003-3978-532X}{\orcidicon}}}
\newcommand\orcidhenrique{{\href{https://orcid.org/0009-0006-7929-6570}{\orcidicon}}}

\newcommand{\orcidicon}{%
	\begin{tikzpicture}
	\draw[lime, fill=lime] (0,0)
		circle [radius=0.16]
		node[white] {{\fontfamily{qag}\selectfont \tiny ID}};
	\draw[white, fill=white] (-0.0625,0.095)
		circle [radius=0.007];
	\end{tikzpicture}	\hspace{-2mm}
}

\usepackage{booktabs}

\begin{document}

\title{Coherent quantum hairy black holes from  gravitational decoupling: regularity, geodesics, and scalar ringdown}

\author{Henrique \textsc{Navarro}\orcidhenrique}
\email{henrique.navarro@ufabc.edu.br}
\affiliation{Federal University of ABC, Center of Physics, Santo Andr\'e, S\~ao Paulo, 09580-210, Brazil}

\author{Roldao \textsc{da Rocha}\orcidroldao\!\!}
\email{roldao.rocha@ufabc.edu.br}
\affiliation{Federal University of ABC, Center of Mathematics, Santo Andr\'e, S\~ao Paulo, 09580-210, Brazil}

\begin{abstract}
We construct a coherent-state quantum extension of gravitational-decoupling (GD) hairy black holes, in which the classical spacetime geometry emerges as the mean-field limit of a finite graviton condensate, while quantum fluctuations provide a natural short-distance regulator. The coherent quantum GD hairy black hole metric is obtained by Gaussian smearing of the gravitational potential, with an effective width encoding the size of the quantum core. The resulting geometry is free of curvature singularities over appropriate parameter ranges and exhibits a modified horizon structure. 
We also investigate geodesic motion in the coherent quantum GD hairy spacetime and find significant deviations from the classical Reissner--Nordström (RN) geometry. In particular, the photon ring, critical impact parameter, and light deflection are modified by the combined effects of quantum corrections and GD hair, providing potential strong-field tests of deviations from general relativity. \textcolor{black}{Finally, we employ the WKB approximation to show that the coherent quantum GD hairy black hole has a quasinormal mode spectrum that differs from those of both the classical GD hairy and Schwarzschild black holes.}

\end{abstract}

\maketitle

\section{Introduction}
\label{sec1}
The detection of gravitational waves (GWs) by laser-interferometric observatories has advanced the understanding of the strong-field regime of gravity \cite{LIGOScientific:2019fpa}. Emitted during the late stages of binary mergers, GWs enable tests of solutions to the Einstein field equations and their extensions. In particular, the ringdown phase of compact binary coalescences is governed by quasinormal modes (QNMs), which are sensitive to deviations from the Kerr and Reissner--Nordström (RN) geometries. Consequently, GW observations provide a powerful probe of black hole hair and short-range modifications of gravity.  

The gravitational decoupling (GD) approach has emerged as a robust framework for constructing self-gravitating compact stellar configurations and black holes, from seed solutions in general relativity (GR) \cite{Ovalle:2017fgl,Ovalle:2018gic}. This technique naturally induces anisotropic matter distributions and allows for analytical solutions of the Einstein field equations through the introduction of additional sources in the energy-momentum tensor, encoding tidal and gauge charges, scalar hair, or any contributions from extended theories of gravity 
~\cite{Yousaf:2024tes,Contreras:2021yxe,Leon:2023nbj,Ramos:2021drk, Sharif:2023ecm, Morales:2018urp, Panotopoulos:2018law, Singh:2019ktp}.  
Relativistic nuclear theory indicates that stellar interiors at high densities develop pressure anisotropies \cite{Gavassino:2023xkt,Berkimbayev:2025ail}, which are naturally accommodated by the  GD method, where the additional gravitational sector induces effective radial and tangential pressures that differ even when starting from isotropic seed solutions
~\cite{Heras:2018cpz, Torres:2019mee, Hensh:2019rtb, Maurya:2023uiy, Iqbal:2025xqf, Tello-Ortiz:2023poi,Khatoon:2025spa,Zubair:2026znq}.  
The GD approach plays a prominent role in gauge/gravity correspondences, accounting for black hole solutions in the infrared (IR) regime \cite{daRocha:2017cxu,Meert:2020sqv,Casadio:2017sze,daRocha:2012pt}, including holographic entanglement entropy approaches \cite{daRocha:2020gee}.  GD black holes have been extensively investigated through gravitational lensing analyses \cite{Cavalcanti:2016mbe,Wang:2025hzu,Liang:2024xif}, redshift measurements \cite{Gabbanelli:2018bhs}, and studies within analog gravity  \cite{Li:2022hkq}.
An important recent advance is the emergence of black hole solutions with gravitational decoupling (GD) hair \cite{Ovalle:2020kpd,Rincon:2019jal,Avalos:2023ywb,Albalahi:2024vpy,Sharif:2025oeu,Andrade:2025xxl,Prihadi:2025rwg,dePaiva:2025eux}. These configurations have also been explored in dark matter phenomenology \cite{Maurya:2024zao,Almatroud:2025xqq,Maurya:2025kto}. GD hairy black holes possess extra degrees of freedom that are not related to quasilocal conserved charges, and the study of such primary hair contributes meaningfully to the consistent analytical modeling of compact objects and black holes within GR and quantum gravity frameworks. Extensions of GD hairy black holes to asymptotically AdS spacetimes have also been reported \cite{Zhang:2022niv,Lin:2024ubg}. From an observational perspective, their QNM mode spectra and associated signatures have been investigated in Refs.~\cite{Guimaraes:2025jsh,Cavalcanti:2022cga,Ribeiro:2025ohn,Yang:2022ifo,Avalos:2023jeh,Al-Badawi:2024iax,Tello-Ortiz:2024mqg}. 
The thermodynamic stability of GD hairy black holes has been analyzed in Refs.~\cite{Ditta:2023arv,Mansour:2024mdg,Mahapatra:2022xea,Misyura:2024fho,Astefanesei:2024wfj}, while possible astrophysical accretion scenarios were proposed in Ref.~\cite{Rehman:2023eor}.

The nature of spacetime at trans-Planckian scales remains one of the central open problems in gravitational physics. Although classical solutions of GR accurately describe macroscopic phenomena, they generically predict curvature singularities, signaling the breakdown of the theory \cite{Casadio:2021gdf}. Black holes, in particular, suggest that quantum effects must modify the classical description in the high-curvature regime. A relevant framework addressing this issue is provided by the corpuscular, or coherent-state, approach to gravity, in which classical fields emerge as macroscopic quantum states of many weakly interacting quanta \cite{Casadio:2017cdv,Casadio:2021eio,Muck:2016stv}. The gravitational field can be described as a coherent state of gravitons, and black holes correspond to condensates of\footnote{Hereon $M_\textsc{p}$ denotes the Planck mass, and $\ell_\textsc{p}$ is the Planck scale.} $N \sim M^2/M_\textsc{p}^2$ soft gravitons whose collective dynamics reproduce the classical geometry in a mean-field limit \cite{Casadio:2015bna,Giusti:2019wdx}. Quantum effects arise from the finite occupation number and manifest themselves as $1/N$ corrections, which become relevant near the would-be singularity or for sufficiently small black holes \cite{Casadio:2016zpl}. This viewpoint is formalized in the quantum $N$-portrait of black holes \cite{Dvali:2011aa}, which provides a microscopic interpretation of black hole entropy, evaporation, and deviations from exact thermal behavior, predicting the existence of quantum hair \cite{Dvali:2012rt}. Within this framework, the coherent state approach to quantum black holes offers a straightforward quantum-mechanical implementation of these ideas, without aiming to constitute an ultraviolet-complete theory of quantum gravity, while capturing leading quantum corrections to the classical black hole geometry. Collective effects of the graviton condensate were further shown to modify semiclassical expectations near the horizon \cite{Flassig:2012re}. These effects were scrutinized in the context of the GD setup \cite{Casadio:2016aum,Fernandes-Silva:2019fez}. A thorough quantum field-theoretical approach to graviton bound states is reported in Refs. \cite{Hofmann:2014jya,Casadio:2013hja}, whereas Refs. \cite{Battista:2023iyu} discusses quantum geometries in the context of effective field theory models of gravity.

Related developments were pursued within effective quantum-geometry approaches, in which a smeared structure replaces a sharply defined horizon. The horizon properties of quantum black holes were studied in Ref. \cite{Casadio:2015jha}. They have since been extended to increasingly realistic quantum-corrected Schwarzschild and RN  geometries, as well as rotating quantum black holes \cite{Cadoni:2018dnd}. The role of quantum hair and its contribution to entropy was investigated in Ref. \cite{Feng:2024bsx}, and quantum effects during gravitational collapse were analyzed in Ref. \cite{Calmet:2023met}. Black holes with quantum matter cores were proposed in Refs. \cite{Casadio:2024lzd,Casadio:2023pmh}, and the construction of coherent quantum black holes carrying electric charge was accomplished, e.g., in Refs. \cite{Feng:2025nai,Antonelli:2025mcv,Meert:2025cct}. Therefore, GD hairy black holes can provide a natural arena in which quantum gravitational effects can manifest at horizon scales while remaining consistent with classical behavior at large distances.

An important consequence of the corpuscular or coherent-state picture is the emergence of an intrinsic length scale associated with the spread of the graviton wavefunction. Since a coherent state cannot support arbitrarily short wavelengths, the resulting spacetime geometry is effectively smeared at small scales, providing a natural mechanism for regularizing curvature divergences \cite{Urmanov:2024qai}. Several recent works have implemented this idea by introducing Gaussian or other smearing functions in momentum space, yielding quantum-corrected black hole metrics that are nonsingular and exhibit modified horizon structures \cite{Cirilo-Lombardo:2023cxb}.

\textcolor{black}{A fundamental aspect of the coherent-state description consists of the emergence of the classical metric
as the semiclassical (mean-field) limit of the gravitational field,
defined as the expectation value of the metric operator over a macroscopic
coherent state composed of a large number of soft graviton quanta. In the effective approach to be adopted here, the finite spatial
extent of the graviton wavefunction is modeled through a Gaussian profile,
representing the characteristic spread of the condensate
\cite{Casadio:2022ndh,Casadio:2024whh}. Rather than quantizing the metric
directly, quantum effects are incorporated at the level of the gravitational
potential, replacing the point-source contribution by a smeared
distribution. }

\textcolor{black}{In this work, we construct coherent quantum GD hairy black holes by applying the
coherent-state Gaussian smearing procedure to the classical GD solution. In
particular, the mass and charge contributions in the metric function are
replaced by their coherent-state counterparts obtained from a graviton
condensate with finite spatial extension, while the GD deformation sector
responsible for the gravitational hair is consistently incorporated. The
resulting geometry preserves the asymptotic behavior of the classical solution
at large distances, where the graviton condensate behaves effectively as a
point source, while significantly modifying the near-core region and
regularizing the central singularity.}

We then analyze the resulting spacetime in detail, computing the effective
stress--energy tensor, Misner--Sharp mass, and the associated energy density
and pressures. We further study the horizon structure and causal properties,
as well as the behavior of curvature invariants near the origin, identifying
conditions under which the geometry is regular. Finally, we examine geodesic
motion and optical observables, highlighting the combined effects of quantum
corrections and GD hair.

This paper is organized as follows. Section \ref{sec2} reviews the classical GD hairy black hole, exploring the role of the hair parameter in deforming the RN geometry. Section \ref{sec:QCRN} introduces the coherent quantum framework with Gaussian regularization, derives the quantum-corrected RN spacetime, and extends it to the GD hairy case, showing that the smearing scale resolves the singularity while preserving large-scale behavior. Section \ref{sec:effective_source} is devoted to obtaining the effective stress–energy tensor and the Misner--Sharp mass, delving into  quantum effects and the influence of the hairy parameter on the ADM mass. The horizon structure and causal properties are also analyzed. Section \ref{sec:regularity} identifies regularity conditions at the origin. Section \ref{geo1} studies geodesics, photon rings, the critical impact parameter, and light deflection, emphasizing deviations from the classical geometry. Section \ref{sec: quasinormal modes} investigates the geometry  QNMs when submitted to scalar perturbations, by employing the WKB approximation to calculate the oscillation frequencies. We show how the real and imaginary frequencies are affected by both the quantum corrections and the GD hair charges, and verify relevant deviations from the classical GD hairy black holes and, in particular, from the RN geometry. Finally, Section \ref{sec6} presents conclusions and outlines possible extensions regarding applications of coherent quantum GD hairy black holes.

\section{GD hairy black holes}
\label{sec2}

In this section,  the GD framework is briefly reviewed and applied to the construction of GD hairy black hole solutions sourced by an additional gravitational sector  \cite{Ovalle:2020kpd}. One considers the Einstein field equations\footnote{We use units with
$c=1$ and $\kappa^2=8\pi G_\textsc{n}$, where $G_\textsc{n}$ is the Newton gravitational constant.}
\begin{equation}
\label{corr2}
R_{\mu\nu}-\frac12R g_{\mu\nu}
=
\kappa^2\left(\scalebox{0.97}{$T$\!}_{\mu\nu}+\mathcal{{\scalebox{0.97}{$\uptheta$}}}_{\mu\nu}\right)\ ,
\end{equation}
with the total energy-momentum tensor composed by a known seed sector $\scalebox{0.97}{$T$\!}_{\mu\nu}$ and an additional
source $\mathcal{{\scalebox{0.97}{$\uptheta$}}}_{\mu\nu}$ \cite{Ovalle:2017fgl,Ovalle:2018gic}. 
Since the Einstein tensor obeys the Bianchi identity, the total source must satisfy
\begin{equation}
\nabla^\mu\left(\scalebox{0.97}{$T$\!}_{\mu\nu}+\mathcal{{\scalebox{0.97}{$\uptheta$}}}_{\mu\nu}\right)=0\ .
\label{csv00}
\end{equation}
For a static and spherically symmetric geometry,
\begin{equation}
\label{metric}
ds^2
=
e^{\upnu(r)}dt^2-e^{\uplambda(r)}dr^2-r^2 d\Omega^2\ ,
\end{equation}
the Einstein field equations~\eqref{corr2} can be written as (the symbol $A^\prime$ indicates the derivative of a function $A(r)$ with respect to  $r$):
\begin{subequations}
\begin{eqnarray}
\label{efe01}
\kappa^2(T_0^{\; 0}+\mathcal{{\scalebox{0.97}{$\uptheta$}}}_0^{\; 0})
&=&
\frac1{r^2}+\frac{e^{-\uplambda}}{r}\!\left({\uplambda'}-\frac1{r}\right),
\\
\label{efe02}
\kappa^2(T_1^{\; 1}+\mathcal{{\scalebox{0.97}{$\uptheta$}}}_1^{\; 1})
&=&
\textcolor{black}{\frac{1}{r^2}-}{e^{-\uplambda}}\!\left(\frac{\upnu'}{r}+\frac1{r^2}\right),
\\
\label{efe03}
\kappa^2(T_2^{\; 2}+\mathcal{{\scalebox{0.97}{$\uptheta$}}}_2^{\; 2})
&=&
-\frac{e^{-\uplambda}}{4}\!
\left(2\upnu''+\upnu'^2-\uplambda'\upnu'
  +2\frac{\upnu'-\uplambda'}r\right).
\end{eqnarray}
\end{subequations}
The anisotropy term is then defined as $
\Pi\equiv  p_\textsc{t}- p_\textsc{r}$, 
with associated effective radial pressure
\begin{equation}
{p}_\textsc{r}
=
-T_1^{\; 1}
-\mathcal{{\scalebox{0.97}{$\uptheta$}}}_1^{\; 1}
\ ,
\label{erp1}
\end{equation} effective tangential pressure defined by 
\begin{equation}
{p}_\textsc{t}
=
-T_2^{\; 2}
-\mathcal{{\scalebox{0.97}{$\uptheta$}}}_2^{\; 2},
\label{erp2}
\end{equation} 
and effective energy density given as 
\begin{equation}
{\rho}
=
T_0^{\; 0}
+
\mathcal{{\scalebox{0.97}{$\uptheta$}}}_0^{\; 0}
\ ,
\label{erp3}
\end{equation}
Therefore, for effective sources generated by the GD procedure, a negative effective radial pressure is required when the total energy density is positive. The condition
\begin{equation}
 p_\textsc{r} = -  \rho 
\label{constr1}
\end{equation}
 is necessary to preserve the existence of a regular Killing horizon and to ensure the GD decoupling. Indeed,  \textcolor{black}{taking the difference between the temporal component \eqref{efe01} and the radial component (\ref{efe02}) of the 
 Einstein field  equations yields} \cite{Ovalle:2020kpd}
\beq\label{eq99}
\kappa^2(\rho+ p_\textsc{r})
=
\frac{e^{-\uplambda}}{r}\left(\upnu'+\uplambda'\right).
\eeq
At the horizon $r=r_h$, where $e^{\upnu(r_h)}=0=e^{-\uplambda(r_h)}$, regularity requires Eq. (\ref{eq99}) be satisfied only if Eq. (\ref{constr1}) holds  \cite{Ovalle:2020kpd}. Near the horizon, the GD-induced sector behaves like a vacuum-like fluid with equation of state (\ref{constr1}), effectively polarizing the gravitational vacuum and generating anisotropic stresses that deform the black hole geometry \cite{Ovalle:2020kpd}.

In the context of the GD approach, let $\upxi(r)$ and $\upchi(r)$ compose,  respectively, the temporal and radial components of the metric for the seed solution as 
\beq\label{pfmetric}
ds^2 = e^{\upxi(r)} dt^2 - e^{\upchi(r)} dr^2 - r^2 d\Omega^2,
\eeq describing the original spacetime geometry before any additional source or deformation is  introduced. The standard GR relation
\begin{equation}
\label{seed0}
e^{-\upchi(r)}
=
1-\frac{\kappa^2}{r}\!\int_0^r d\mathsf{r}\,\mathsf{r}^2 T_0^{\; 0}(\mathsf{r})
\equiv
1-\frac{2m(r)}r
\end{equation}
defines the Misner--Sharp mass $m(r)$, interpolating between the local energy content at finite radius and the total gravitational mass measured at infinity. 

Any known seed solution of the Einstein field equations is modified to include the effects of the additional source $\mathcal{{\scalebox{0.97}{$\uptheta$}}}_{\mu\nu}$. 
GD deformations -- described in the metric (\ref{metric}) -- of the seed metric (\ref{pfmetric}) are introduced to guarantee the separability of the Einstein field equations, as 
\begin{subequations}
\beq\label{gd1}
\upnu(r)&=&\upxi(r)+\alpha g(r),\\
\label{gd2}
e^{-\uplambda(r)}&=&e^{-\upchi(r)}+\alpha h(r),
\eeq 
\end{subequations}
where $\alpha$ controls the strength of the GD deformation. Eqs. (\ref{gd1}, \ref{gd2}) ensure that the Einstein field equations split into two independent systems, one corresponding to the original seed solution and the other associated with the additional gravitational sector. 
Requiring that the exterior solution remains asymptotically flat and matches the Schwarzschild geometry yields the constraint \cite{Ovalle:2020kpd} \begin{equation}
\label{fg}
\alpha h(r)
=
\left(1-\frac{2M}{r}\right)\left(e^{\alpha g(r)}-1\right). 
\end{equation} 
In this way, $g(r)$ and $h(r)$ in Eqs. (\ref{gd1}, \ref{gd2}) introduce the hair generated by the additional source, allowing one to systematically explore deviations from standard GR while maintaining a consistent exterior GD solution. 
Therefore the line element~\eqref{metric} yields 
\begin{eqnarray}
\label{hairyBH}
ds^{2}
&=&
\left(1-\frac{2M}{r}\right)
e^{\alpha g(r)}
dt^{2}
-\left(1-\frac{2M}{r}\right)^{-1}
e^{-\alpha g(r)}
dr^2
-r^{2}d\Omega^2
\ .
\end{eqnarray} Besides, the seed energy-momentum tensor  is conserved with respect to the seed metric components in \eqref{pfmetric}:
\begin{equation}
\label{pfcon}
\nabla^{(\upxi,\upchi)}_\mu \left(\scalebox{0.97}{$T$\!}^{\,\mu}_{\;\,\,\nu}+\mathcal{{\scalebox{0.97}{$\uptheta$}}}^{\,\mu}_{\;\,\,\nu}\right)=0.
\end{equation}
Substituting the GD deformations (\ref{gd1}, \ref{gd2}) into the Einstein field equations yields two
independent systems.  The first one comprises the seed sector, represented by 
\begin{subequations}
\begin{eqnarray}
\label{ec1pf}
\kappa^2 T_0^{\; 0}
&=&
\frac1{r^2}-\frac{e^{-\upchi}}{r}\!\left(\frac1{r}-{\upchi'}\right),\\
\label{ec2pf}
\kappa^2 T_1^{\; 1}
&=&
\frac1{r^2}-\frac{e^{-\upchi}}{r}\!\left(\frac1{r}+{\upxi'}\right),\\
\label{ec3pf}
\kappa^2 T_2^{\; 2}
&=&
-\frac{e^{-\upchi}}4
\left(2\upxi''+\upxi'^2-\upchi'\upxi'+2\frac{\upxi'-\upchi'}r\right).
\end{eqnarray}
\end{subequations}
The second sector consists of the quasi-Einstein system for the additional source $\mathcal{{\scalebox{0.97}{$\uptheta$}}}_{\mu\nu}$  \cite{Ovalle:2020kpd,Zubair:2023cvu, Bamba:2023wok}:
\begin{subequations}
\begin{eqnarray}
\label{ec1d}
\kappa^2\mathcal{{\scalebox{0.97}{$\uptheta$}}}_0^{\; 0}
&=&
-\frac{\alpha}{r}\left(\frac{h}{r}+h'\right),
\\
\label{ec2d}
\kappa^2\mathcal{{\scalebox{0.97}{$\uptheta$}}}_1^{\; 1}+\alpha \mathfrak{z}_1
&=&
-\frac{\alpha h}{r}\!\left(\frac1{r}+\upnu'\right),
\\
\label{ec3d}
\kappa^2\mathcal{{\scalebox{0.97}{$\uptheta$}}}_2^{\; 2}+\alpha \mathfrak{z}_2
&=&
-\frac{\alpha h}{4}
\left(2\upnu''+\upnu'^2+2\frac{\upnu'}r\right)
-\frac{\alpha h'}4\!\left(\upnu'+\frac2r\right),
\end{eqnarray}
\end{subequations}
with
\begin{subequations}
\begin{eqnarray}
\label{Z1}
\mathfrak{z}_1&=&\frac{e^{-\upchi} g'}r,\\
\label{Z2}
\mathfrak{z}_2&=&\frac{e^{-\upchi}}4\!\left(2g''+g'^2+\frac{2g'}r+2\upxi' g'-\upchi' g'\right).
\end{eqnarray}
\end{subequations}
Clearly $\displaystyle\lim_{\alpha\to0}\mathcal{{\scalebox{0.97}{$\uptheta$}}}_{\mu\nu}=0$, and for $g=0$ the quasi-Einstein system (\ref{ec1d}) --  (\ref{ec1d}) reduces to the minimal geometric deformation in Ref.~\cite{Ovalle:2017fgl}.

The conservation equation~\eqref{csv00} becomes
\begin{equation}
\nabla_\sigma \scalebox{0.97}{$T$\!}^{\;\sigma}{}_{\nu}
=
\nabla^{(\upxi,\upchi)}_\sigma \scalebox{0.97}{$T$\!}^{\;\sigma}{}_{\nu}
-\alpha\frac{g'}2(T_0^{\; 0}-T_1^{\; 1})\delta^1_\nu ,
\label{divs}
\end{equation}
where the first term on the right-hand side of \eqref{divs}  vanishes due to the conservation law in Eq.~\eqref{pfcon}. 
Using the full metric~\eqref{metric}, the total conservation law then gives
\begin{eqnarray}
\!\!\!\!\!\!\!\!\!\!\!\!\!\!\!\!\!\!\!\!0
&=&
\nabla_\sigma \scalebox{0.97}{$T$\!}^{\;\sigma}{}_{\nu}
+\nabla_\sigma\mathcal{{\scalebox{0.97}{$\uptheta$}}}^{\sigma}{}_{\nu}
=
\left(\mathcal{{\scalebox{0.97}{$\uptheta$}}}_1^{\; 1}\right)'
-\frac{\upnu'}2(\mathcal{{\scalebox{0.97}{$\uptheta$}}}_0^{\; 0}-\mathcal{{\scalebox{0.97}{$\uptheta$}}}_1^{\; 1})
-\frac2r(\mathcal{{\scalebox{0.97}{$\uptheta$}}}_2^{\; 2}-\mathcal{{\scalebox{0.97}{$\uptheta$}}}_1^{\; 1})
-\alpha\frac{g'}2 (T_0^{\; 0}-T_1^{\; 1}),
\end{eqnarray}
which is a linear combination of Eqs.~\eqref{ec1d} - \eqref{ec3d}.  
Thus the seed sector $\scalebox{0.97}{$T$\!}_{\mu\nu}$ and the additional source ${\scalebox{0.97}{$\uptheta$}}_{\mu\nu}$ are fully decoupled by the
GD.  This protocol was extended in the context of Lovelock gravity \cite{Estrada:2019aeh}, Brans-Dicke gravity  \cite{Sharif:2020lbt}, Riemann--Cartan spacetimes, and Einstein--Maxwell-scalar gravities  \cite{Mahapatra:2020wym,Pradhan:2025dno,Priyadarshinee:2021rch,Naseer:2025fwm}. 

The dominant energy condition (DEC) requires
\begin{subequations}
\beq
\rho&\ge&| p_\textsc{r}|,\label{dec-basic0}\\ 
\rho&\ge&| p_\textsc{t}|.
\label{nt}
\eeq
\end{subequations}
Using Eq. (\ref{constr1}) yields the only nontrivial condition (\ref{nt}) which, using Eqs.~(\ref{erp2}, \ref{erp3}), is equivalent to \cite{Ovalle:2020kpd}  the two inequalities:
\begin{eqnarray}
{\scalebox{0.97}{$\uptheta$}}_0^{\; 0}+{\scalebox{0.97}{$\uptheta$}}_2^{\; 2}\ge0\ ,\qquad\qquad
{\scalebox{0.97}{$\uptheta$}}_0^{\; 0}-{\scalebox{0.97}{$\uptheta$}}_2^{\; 2}\ge0\ .
\label{dec-b}
\end{eqnarray}
Defining the radial function
\beq\label{kgr}
k(r)\equiv e^{\alpha g(r)},
\eeq and using Eqs.~\eqref{ec1d} and \eqref{ec3d}, the first inequality in   (\ref{dec-b}) yields 
\begin{eqnarray}
{\cal H}_1(r)&\equiv&r(r-2G_\textsc{n} M)k''(r)+4(r-G_\textsc{n} M)k'(r)+2k(r)-2\le0\ ,
\label{H1}
\end{eqnarray}
One looks for deformations that satisfy the inequality (\ref{H1}) near the horizon $r\approx2 G_\textsc{n}M$ and at the outer geometry   
$r\gg M$.

The differential inequality \eqref{H1} admits a family of admissible deformations compatible with the DEC. To proceed analytically, one focuses on solutions that saturate the inequality \eqref{H1}, with a source \cite{Ovalle:2020kpd}
\begin{equation}\label{sfg}
{\cal H}_1(r)=\frac{\alpha}{G_\textsc{n} M}(r-2G_\textsc{n} M)e^{-r/(G_\textsc{n} M)}.
\end{equation}
Therefore, the solution of the resulting equation is given by 
\begin{equation}
k(r)=
1-\frac{1}{r-2G_\textsc{n}M}
\left[
\alpha\ell
+\alpha G_\textsc{n} M e^{-r/(G_\textsc{n} M)}
-\frac{G_\textsc{n} Q^2}{r}
\right],
\label{h-dec-sol}
\end{equation}
where $\ell$ and $Q$ encode the black hole hair proportional to~$\alpha$.  Substituting the function~\eqref{h-dec-sol} into the metric~\eqref{hairyBH}, using Eq. \eqref{kgr}, finally yields the GD hairy black hole metric coefficients of the metric (\ref{metric}), with Yukawa-type exponential term \cite{Ovalle:2020kpd}: 
\begin{equation}
e^{\upnu(r)}=e^{-\uplambda(r)}
=
1-\frac{2G_\textsc{n} M+\alpha G_\textsc{n}\ell}{r}
+\frac{G_\textsc{n} Q^2}{r^2}
-\frac{\alpha G_\textsc{n} M}{r}e^{-r/(G_\textsc{n}M)},
\label{gdd}
\end{equation}

An important aspect of the GD hairy black hole solution under DEC is that it does not allow for arbitrary values of the hair charges $\ell$ and $Q$, since it requires a horizon radius $r_h \geq 2 G_\textsc{n} M$. Ref. \cite{Ovalle:2020kpd} shows that these charges must present the following lower bounds:
\begin{equation}
    Q^2 \geq \frac{4 \alpha M^2}{e^2}, \qquad\qquad \qquad \ell \geq \frac{M}{e^2}. \label{eq: lower bounds over l}
\end{equation}

\section{Coherent quantum GD hairy black holes}
\label{sec:QCRN}

For static, spherically symmetric and asymptotically flat configurations,
the line element can be written as
\begin{equation}
ds^2 = -f(r)\,dt^2 + \frac{dr^2}{f(r)} + r^2 d\Omega^2,
\label{eq:metric_ansatz}
\end{equation}
with
\begin{eqnarray}\label{mf2}
    f(r) \equiv 1 + 2V(r),
\end{eqnarray}
where $r$ denotes the areal radius, and the horizons are determined by the
solutions of $f(r)=0$. 
We will show that the GD hairy solution (\ref{metric}, \ref{gdd}) represents a classical spacetime geometry that can be interpreted as the expectation value of a coherent quantum GD hairy metric operator evaluated on a suitable coherent state built upon the Minkowski vacuum ~\cite{Casadio:2016zpl,Casadio:2017cdv}. 
This approach allows one to encode the gravitational field in an effective potential that appears in the metric. 
A static and spherically symmetric gravitational potential $V(r)$ can be identified with the expectation
value of a canonically normalized free massless scalar field operator
$\Phi$ evaluated over a coherent state $|\,g\,\rangle$ \cite{Casadio:2021eio},
\begin{equation}
\langle\, g\, | \Phi(t,r) | \,g\, \rangle = V(r) .
\label{eq:Phi_expectation}
\end{equation}
The scalar field $\Phi$ does not represent a fundamental matter degree
of freedom, but provides an effective description of the collective,
non-perturbative quantum degrees of freedom underlying the classical geometry. The quantization of the scalar field can be implemented by first rescaling the dimensionless potential $V$ as
\begin{equation}
V(r)\mapsto \sqrt{\frac{m_\textsc{p}}{\ell_\textsc{p}}}\, V(r) ,
\end{equation}
so that $\Phi$ satisfies the free massless Klein--Gordon equation,
\begin{equation}
\left( -\partial_t^2 + \nabla^2 \right)\Phi(t,r) = 0,
\label{eq:KG}
\end{equation} whose complete set of solutions is given by 
$
\mathsf{u}_k(t,r) = e^{- i k t} j_0(kr),
$ 
where $j_0(kr) = \displaystyle\frac{\sin(kr)}{kr}$ is the zeroth order spherical Bessel function, 
with $k>0$.
The field operator and its conjugate momentum, respectively, read
\begin{subequations}
\begin{align}\label{gphi}
\Upphi(t,r)
&=
\frac{1}{2\pi^2}\int_0^\infty {k^2 \, dk}
\sqrt{\frac{\hbar}{2k}}
\left[
 a_k \mathsf{u}_k(t,r) +  a_k^\dagger \mathsf{u}_k^*(t,r)
\right], \\
\Uppi(t,r)
&=
\frac{i}{2\pi^2} \int_0^\infty {k^2 \, dk}
\sqrt{\frac{\hbar k}{2}}
\left[
 a_k \mathsf{u}_k(t,r) -  a_k^\dagger \mathsf{u}_k^*(t,r)
\right],
\end{align}
\end{subequations}
where the creation and annihilation operators satisfy
$\displaystyle
[a_k ,  a_p^\dagger]
=
\frac{2\pi^2}{k^2} \delta(k-p) .$ 
Classical field configurations are reproduced by coherent states $|\,g\,\rangle$
defined as eigenstates of the annihilation operators,
\begin{equation}\label{gkk}
 a_k |\,g\,\rangle = g_k e^{ikt} |\,g\,\rangle,
\end{equation}
removing explicit time
dependence on the expectation value, for static potentials.
Writing the Fourier--Bessel decomposition
\begin{equation}
V(r)
=
\frac{1}{2\pi^2}\int_0^\infty k^2 
\tilde V(k)\, j_0(kr)\, dk  ,
\label{eq:VBessel}
\end{equation}
one finds in Eq. (\ref{gkk})
\begin{equation}
g_k
=
\sqrt{\frac{k}{2}}\,
\frac{\tilde V(k)}{\ell_\textsc{p}} .
\label{eq:gk}
\end{equation}
The coherent state then reads
\begin{equation}
|\,g\,\rangle
=
\frac{e^{-N_\textsc{g}/2}}{2\pi^2}
\exp\!\left[
\int_0^\infty {k^2 }
g_k \, dk a_k^\dagger\,
\right]
|0\rangle ,
\end{equation}
with the total occupation number
\begin{equation}
N_\textsc{g}
=
\frac{1}{2\pi^2}\int_0^\infty k^2 \,g_k^2\, dk .
\label{eq:NG}
\end{equation}
When applied to the RN geometry with $\alpha=0$ in the metric (\ref{metric}, \ref{gdd}), the classical potential cannot be exactly reproduced by a normalizable coherent state due to ultraviolet (UV) divergences associated with high-momentum modes~\cite{Casadio:2022ndh}. This obstruction is resolved by introducing a finite smearing scale $R_{\rm s}$, interpreted as the size of the quantum core, and implementing a Gaussian regulator. The resulting quantum-corrected GD hairy potential remains asymptotically RN at large distances while yielding a regular geometry at short scales, providing a consistent effective description of charged black holes within the coherent-state formalism \cite{Antonelli:2025mcv}.

The scalar field $\Upphi$ in Eq.~(\ref{gphi}) effectively describes the quantum constituents that make classical geometry to emerge. For the RN geometry, the classical gravitational potential reads
\begin{equation}
V_\textsc{rn}(r)
= -\frac{G_\textsc{n} M}{r} + \frac{G_\textsc{n} Q^2}{2r^2},
\label{eq:VRN_classical}
\end{equation}
where $M$ and $Q$ denote the ADM mass and electric charge. Within the coherent-state setup, the potential (\ref{eq:VRN_classical}) cannot be exactly reproduced by a normalizable coherent state of a free massless scalar field due to UV divergences in the occupation number from arbitrarily large-momentum modes \cite{Casadio:2016zpl,Casadio:2017cdv}. In particular, the $r^{-2}$ term yields a slow decay of the Fourier--Bessel transform at large $k$, making the total occupation number divergent and the coherent state non-normalizable \cite{Casadio:2022ndh}.

To cure this issue, one introduces a finite smearing scale $R_{\rm s}$, interpreted as the characteristic size of the quantum core, and adopts a Gaussian regulator rather than a sharp UV cutoff to avoid spurious oscillations in position space. This smoothly suppresses high-momentum modes while preserving the large-distance behavior of the classical solution. For a spherically symmetric configuration, the Gaussian-regularized quantum expectation value of the potential reads
\begin{equation}
V_\textsc{rn}^{q}(r)
= \frac1{2\pi^2}\int_0^\infty {k^2}\,
\tilde V_\textsc{rn}(k)\,
e^{-k^2 R_{\rm s}^2/4}\,
j_0(kr)\,dk,
\label{eq:VRNq_def}
\end{equation}
where the Fourier--Bessel transform of the classical RN potential \eqref{eq:VRN_classical} is given by 
\begin{equation}
\tilde V_\textsc{rn}(k)
= -\frac{4\pi G_\textsc{n} M}{k^2}
+ \frac{\pi^2 G_\textsc{n} Q^2}{k}.
\label{eq:VRN_FT}
\end{equation}
The Gaussian factor $e^{-k^2 R_{\rm s}^2/4}$ in Eq. \eqref{eq:VRNq_def} suppresses large-momentum modes, implementing a minimal-length smearing of width $R_{\rm s}$ in position space. Physically, this replaces point-like mass and charge distributions with Gaussian profiles, removing the UV modes responsible for the classical curvature singularity.

\textcolor{black}{The Gaussian regulator should not be interpreted as an ad hoc regularization prescription, but rather as the momentum-space representation of the underlying coherent graviton state. In the coherent-state framework, the classical gravitational field is reconstructed as the expectation value of a field operator evaluated over a macroscopic condensate of soft gravitons. Consequently, the resulting effective potential is determined by the momentum-space profile of the coherent state itself. Within the corpuscular picture of black holes, the geometry is generated by a bound state containing a large number of weakly interacting gravitons whose collective wavefunction possesses a finite spatial extent. Since such a condensate cannot be localized with arbitrary precision, the source is intrinsically delocalized below a characteristic length scale determined by the finite occupation number $N$, leading to an unavoidable momentum uncertainty. The Gaussian profile naturally encodes this physical dispersion and provides the simplest minimal-uncertainty distribution compatible with a normalizable coherent state. As a result, UV modes are smoothly suppressed due to the finite resolution of the quantum state, rather than by the introduction of an external cutoff.}

\textcolor{black}{From a phenomenological perspective, the smearing scale $R_{\rm s}$ can be interpreted as the effective size of the quantum core of the black hole, determined by the collective binding and coherence properties of the graviton condensate. It plays a role analogous to a coherence length, below which the classical notion of spacetime as a sharply localized geometry ceases to be meaningful. In this sense, the Gaussian profile reflects the loss of localizability of spacetime events at scales comparable to the characteristic wavelength distribution of the constituent gravitons. The regularity of the geometry at short distances therefore emerges as a direct consequence of the underlying quantum state of the gravitational field. Thus, the Gaussian smearing in Eq.~(\ref{eq:VRNq_def}) should be viewed as the effective momentum-space profile of the graviton condensate and as a manifestation of the intrinsic nonlocality associated with the coherent quantum state, rather than as an externally imposed regularization prescription.
}

\textcolor{black}{In addition, in the coherent-state framework, the classical gravitational field is reconstructed as the expectation value of a field operator over a macroscopic graviton condensate, whose momentum-space profile determines the resulting effective potential. In this sense, the smearing function should be understood as encoding the wavepacket structure of the underlying coherent state rather than being an external assumption. The Gaussian profile adopted in this work corresponds to a minimal-uncertainty,  UV-regular distribution, consistent with previous implementations of coherent-state black hole geometries. It ensures a well-defined occupation number and provides a smooth interpolation between the classical regime and the quantum-corrected core.}

Computing the integral (\ref{eq:VRNq_def}) yields
\begin{equation}
V_\textsc{rn}^{q}(r)
= -\frac{G_\textsc{n} M}{r}\,
\erf\!\left(\frac{r}{R_{\rm s}}\right)
+ \frac{G_\textsc{n} Q^2}{r\,R_{\rm s}}\,
\F\!\left(\frac{r}{R_{\rm s}}\right),
\label{eq:VRNq_final}
\end{equation}
where $\erf(x)$ is the error function and
$\displaystyle \F(x)=e^{-x^2}\!\int_0^x e^{t^2}dt$ is the Dawson function.
The Gaussian smearing modifies only the short-distance sector while
leaving the IR behavior intact. For $r\gg R_{\rm s}$,  the classical RN potential can be  recovered.
On the other hand, near the origin, one can expand Eq. (\ref{eq:VRNq_final}), yielding 
\begin{equation}
V_\textsc{rn}^{q}(r)
=
-\frac{2G_\textsc{n}M}{\sqrt{\pi}R_{\rm s}}
+\frac{G_\textsc{n}Q^2}{R_{\rm s}^2}
+\mathcal{O}(r^2).
\end{equation}
The singular $r^{-1}$ and $r^{-2}$ behaviors are therefore replaced by a
finite constant core, implying that the effective gravitational field
vanishes linearly as $r\to0$. The central RN geometry is thus regular,
with curvature invariants remaining finite, and the classical
RN singularity is resolved by the Gaussian smearing \cite{Antonelli:2025mcv}. 
Introducing the definitions 
\begin{equation}
R_{\rm M} \equiv 2G_\textsc{n}M,
\qquad
R_{\rm Q} \equiv \sqrt{G_\textsc{n}}\,Q,
\end{equation}
the quantum-corrected RN metric function becomes \cite{Antonelli:2025mcv}
\begin{equation}
f(r)
= 1
- \frac{R_{\rm M}}{r}\,
\erf\!\left(\frac{r}{R_{\rm s}}\right)
+ \frac{2R_{\rm Q}^2}{r \, R_{}s}\,
\F\!\left(\frac{r}{R_{\rm s}}\right).
\label{eq:f_RN_quantum}
\end{equation}
Expanding around the origin, one finds
\begin{equation}
f(r)
=
1
-\frac{2R_{\rm M}}{\sqrt{\pi}R_{\rm s}}
+\frac{2R_{\rm Q}^2}{R_{\rm s}^2}
+\mathcal{O}(r^2),
\end{equation}
so that $\displaystyle\lim_{r\to0} f(r)$ is finite and the quantum RN spacetime is
regular at the origin. The Gaussian width $R_{\rm s}$ acts moreover as a fundamental scale controlling the transition between
a classical exterior geometry and a nonsingular quantum core.

Now let us consider the GD hairy black hole solution (\ref{gdd}),
whose potential can be split as the sum of an RN part and a hairy component, as 
\begin{equation}
V(r)=V_\textsc{rn}(r)+V_H(r),
\end{equation}
with
\begin{equation}\label{vh111}
{\color{black}
V_H(r)
=-\frac{\alpha G_\textsc{n}}{2r}
\left(
M e^{-r/(G_\textsc{n}M)}+\ell
\right).}
\end{equation}
The exponential term introduces a Yukawa-type correction
characterized by the scale $G_\textsc{n}M$, while the parameter
$\ell$ renormalizes the effective mass, as given in the classical GD hairy metric (\ref{gdd}).
Classically, these terms alter the near-horizon structure, but
retain the short-distance divergence. 

{\color{black}The quantum deformation of the GD hairy black hole is implemented within the coherent-state framework by modeling the gravitational source as a finite-width graviton condensate. In practice, this amounts to replacing the classical point-source mass distribution with a Gaussian-smeared profile characterized by a minimal length scale, which encodes leading finite-$N$ effects of the underlying quantum state. 
The GD deformation sector is then consistently incorporated at the level of the metric function, ensuring that both the standard gravitational contributions and the additional hair terms are modified by the same smearing prescription. This procedure yields a quantum-corrected GD hairy geometry that reduces to the classical solution in the limit when the finite smearing scale,  interpreted as the size of the quantum core, goes to zero, while remaining regular at the origin due to the nonlocal structure induced by the coherent-state construction.}

{\color{black}Applying Gaussian regularization to the quantum hairy part of the potential yields 
\begin{align}
 V_{\rm H}^{\rm q} (r) = \frac{\alpha R_{\rm M}}{8r} e^{R_{\rm s}^2/R_{\rm M}^2} &\left[e^{2r/R_{\rm M}} \, \operatorname{erfc}\left(\frac{R_{\rm s}}{R_{\rm M}} + \frac{r}{R_{\rm s}}\right) - e^{-2r/R_{\rm M}} \, \operatorname{erfc}\left(\frac{R_{\rm s}}{R_{\rm M}} - \frac{r}{R_{\rm s}}\right)\right]\nonumber \\
    &- \frac{\alpha G_{\rm N} \ell}{2r} \, \operatorname{erf} \left(\frac{r}{R_{\rm s}}\right),
\label{eq:VHq_final}
\end{align}
which remains finite as $r\to0$, ensuring that the additional GD hair does not reintroduce a curvature singularity.}
The complete quantum-corrected GD hairy potential can be expressed as 
\begin{equation}\label{mf0}
V_\textsc{gdh}^{q}(r)
= V_\textsc{rn}^{q}(r) + V_H^{q}(r),
\end{equation}
and the metric function (\ref{gdd}) takes the form
\begin{equation}
f(r)=1+2V_\textsc{gdh}^{q}(r).
\label{mf}
\end{equation}
For $\alpha\to0$, the quantum RN geometry can be recovered.
For a nonvanishing hairy parameter $\alpha$, the regular quantum core persists.
The Gaussian smearing 
simultaneously resolves the central singularity and regularizes 
the additional hair contributions, without modifying the
asymptotic RN behavior.

\begin{figure}[h!]
    \includegraphics[width=0.48\linewidth]{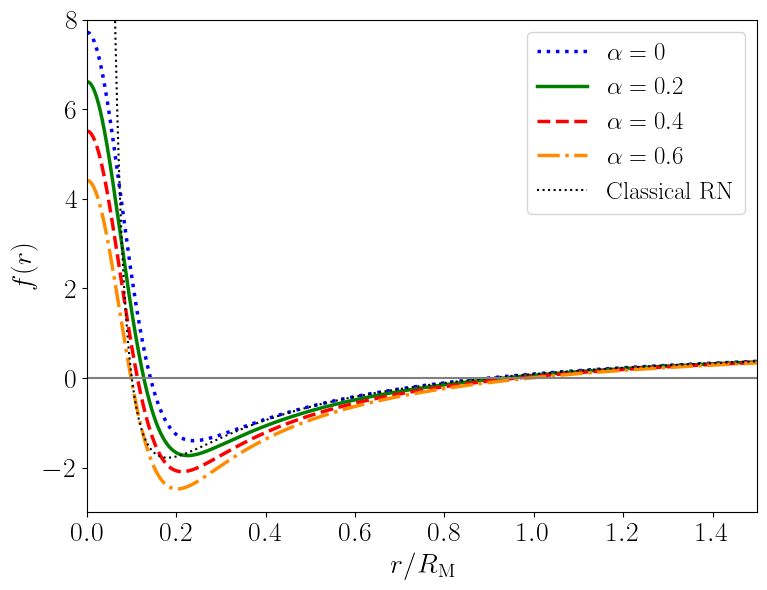}
    \includegraphics[width=0.48\linewidth]{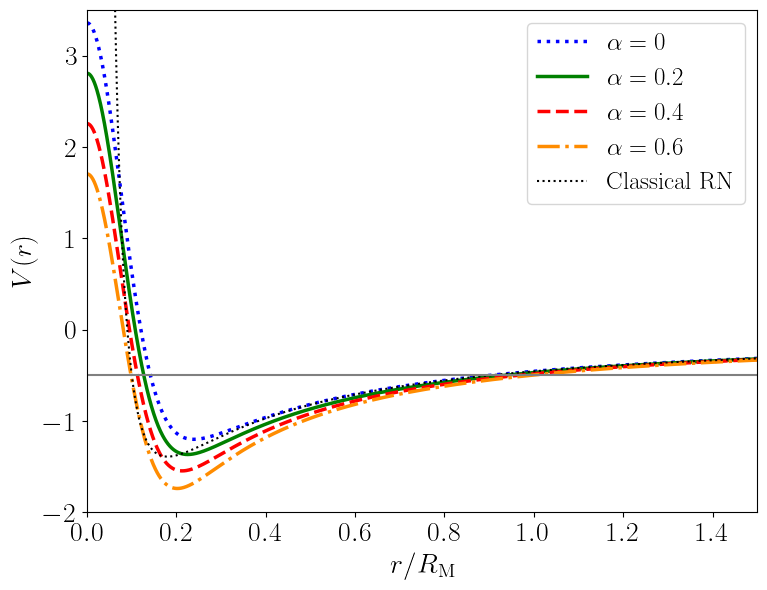}
    \caption{\small {Metric function $f(r)$ (left) and potential function $V_\textsc{gdh}(r)$ (right) for the coherent quantum GD hairy black hole, with $R_{\rm s}/R_{\rm M} = 0.1$ and $R_{\rm Q}/R_{\rm M} = 0.3$.}}
    \label{fig: metric and potential functions for RS=0.05 and RQ=0.1}
\end{figure}

\begin{figure}[h!]    \includegraphics[width=0.48\linewidth]{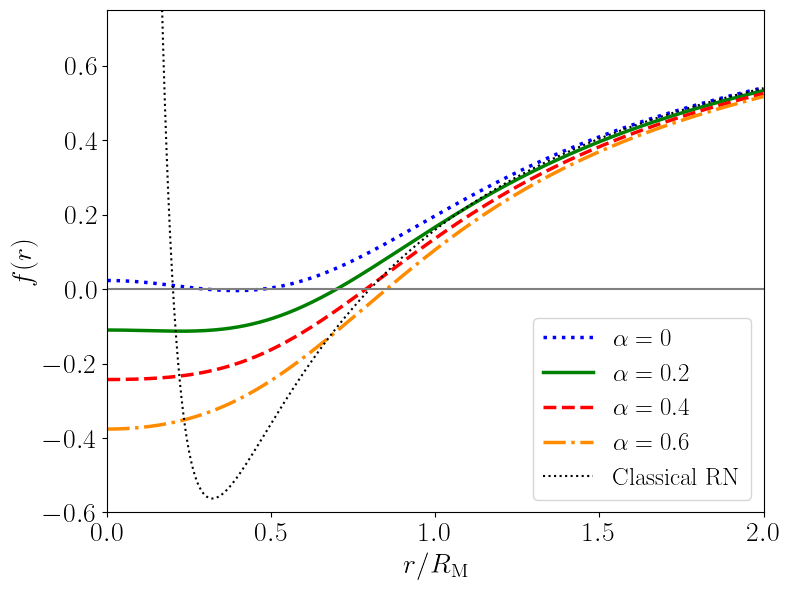}
    \includegraphics[width=0.48\linewidth]{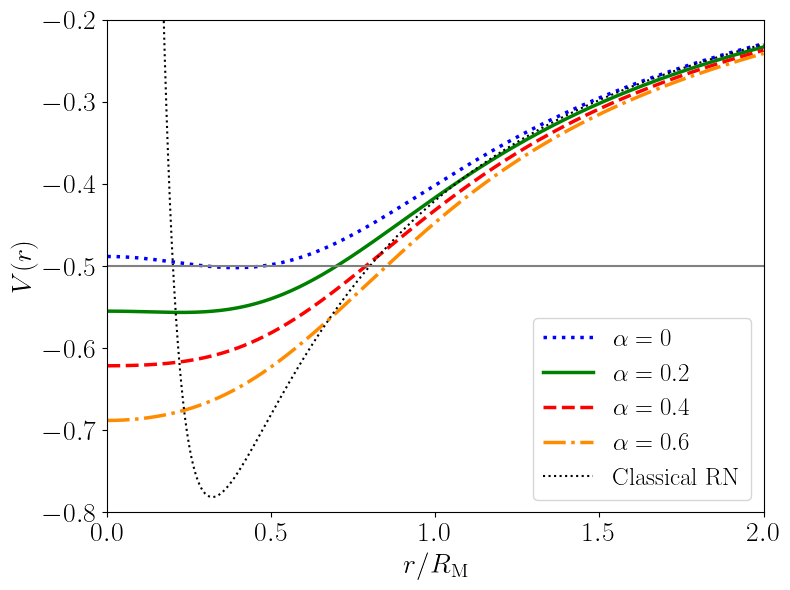}
    \caption{\small {Metric function (left) and potential function (right) for the coherent quantum GD hairy black hole, with $R_{\rm s}/R_{\rm M} = 0.5$ and $R_{\rm Q}/R_{\rm M} = 0.4$.}}
    \label{fig: metric and potential functions for RS=0.5 and RQ=0.4}
\end{figure}

Figs. \ref{fig: metric and potential functions for RS=0.05 and RQ=0.1} and  \ref{fig: metric and potential functions for RS=0.5 and RQ=0.4} present the radial profile of the coherent quantum GD metric function $f(r)$ and their corresponding potential $V^q_\textsc{gdh}(r)$ for different values of the hairy parameter $\alpha$ and representative choices of $R_{\rm s}/R_{\rm M}$ and $R_{\rm Q}/R_{\rm M}$. {\color{black}For both cases depicted, we can see that the metric functions $f(r)$, as well as the associated potential $V(r)$, remain finite at $r=0$ in contrast with the representative Classical RN case. This is a consequence of the quantum regularization procedure that introduces the quantum core of finite size $R_{\rm s}$, and it is independent of the hairy charge $\alpha$, as can be seen by the black dotted lines in Figs \ref{fig: metric and potential functions for RS=0.05 and RQ=0.1} and \ref{fig: metric and potential functions for RS=0.5 and RQ=0.4}, which represent the purely quantum--corrected hairless geometry. As for the $\alpha \neq 0$ configurations, when $\alpha$ increases, the value of $f(r)$ in the inner region decreases. For the $R_{\rm s}/R_{\rm M} = 0.1$ and $R_{\rm Q}/R_{\rm M} = 0.3$ configuration (Fig. \ref{fig: metric and potential functions for RS=0.05 and RQ=0.1}), the qualitative decay pattern at small $r/R_{\rm M}$ is preserved, while for the $R_{\rm s}/R_{\rm M} = 0.5$ and $R_{\rm Q}/R_{\rm M} = 0.4$ configuration (Fig. \ref{fig: metric and potential functions for RS=0.5 and RQ=0.4}), the $f(r)$ behavior in the inner region is significantly altered in comparison with the classical configuration. This shows that the introduction of the quantum core and the hair charges can provide physically meaningful modifications to the black hole geometry. In all cases, the solutions converge to a common asymptotic behavior at large radii, indicating that variations in $\alpha$, $R_{\rm s}$, and $R_{\rm Q}$ predominantly affect the near-core structure of the geometry, while the classical GR solutions are recovered for larger $r$.}

A more illustrative view of our results is depicted in Fig. \ref{fig:f(r, alpha)}, for some values of $R_{\rm s}/R_{\rm M}$ and $R_{\rm Q}/R_{\rm M}$.
\begin{figure}[h!]
    \centering
    \subfigure[$\;\;R_{\rm s}/R_{\rm M} = 0.2$ and $R_{\rm Q}/R_{\rm M}=0.4.$]{
        \includegraphics[width=0.45\textwidth]{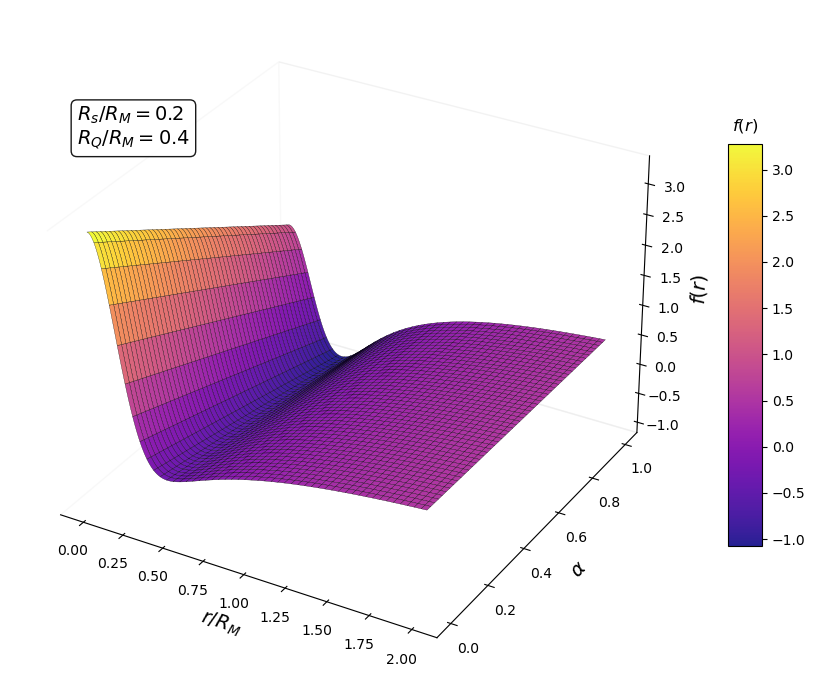}
        \label{fig:a0.2}
    }
    \hfill
    \subfigure[$\;\;R_{\rm s}/R_{\rm M} = 0.3$ and $R_{\rm Q}/R_{\rm M}=0.45.$]{
        \includegraphics[width=0.45\textwidth]{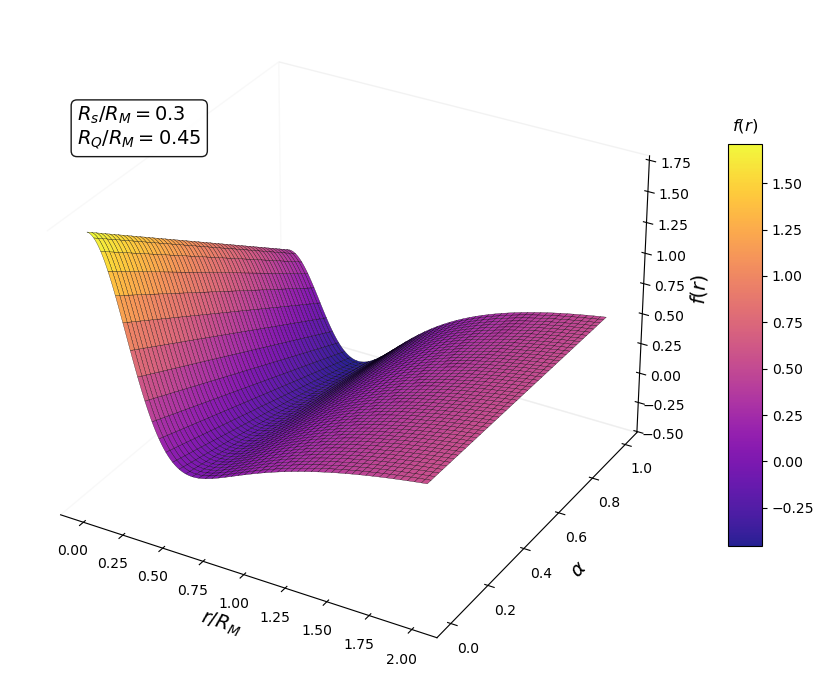}
        \label{fig:a0.3}
    }
    
    \vspace{0.5cm} 
    
    \subfigure[$\;\;R_{\rm s}/R_{\rm M} = 0.5$ and $R_{\rm Q}/R_{\rm M}=0.1.$]{
        \includegraphics[width=0.45\textwidth]{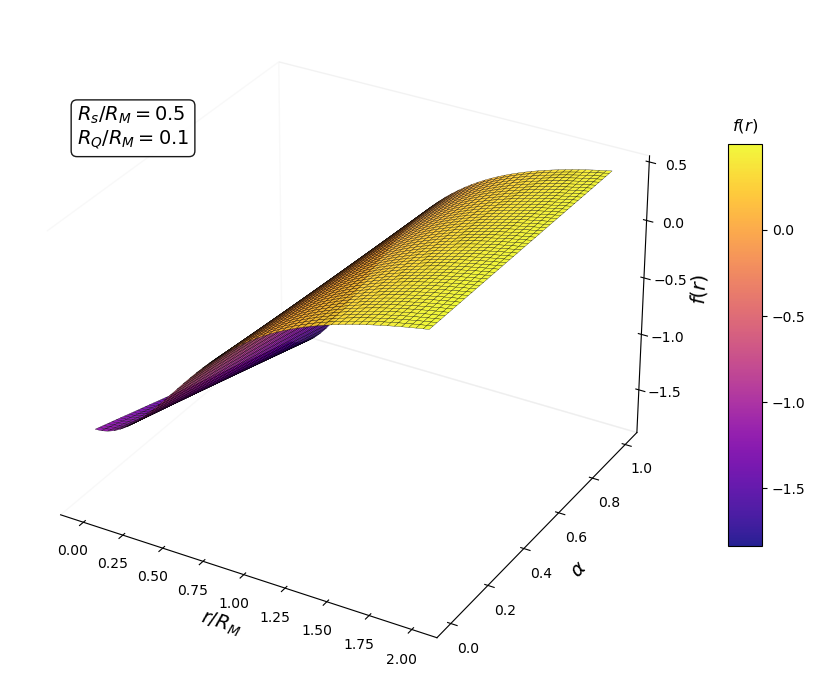}
        \label{fig:a0.5}
    }
    \hfill
    \subfigure[$\;\;R_{\rm s}/R_{\rm M} = 0.7$ and $R_{\rm Q}/R_{\rm M}=0.5.$]{
        \includegraphics[width=0.45\textwidth]{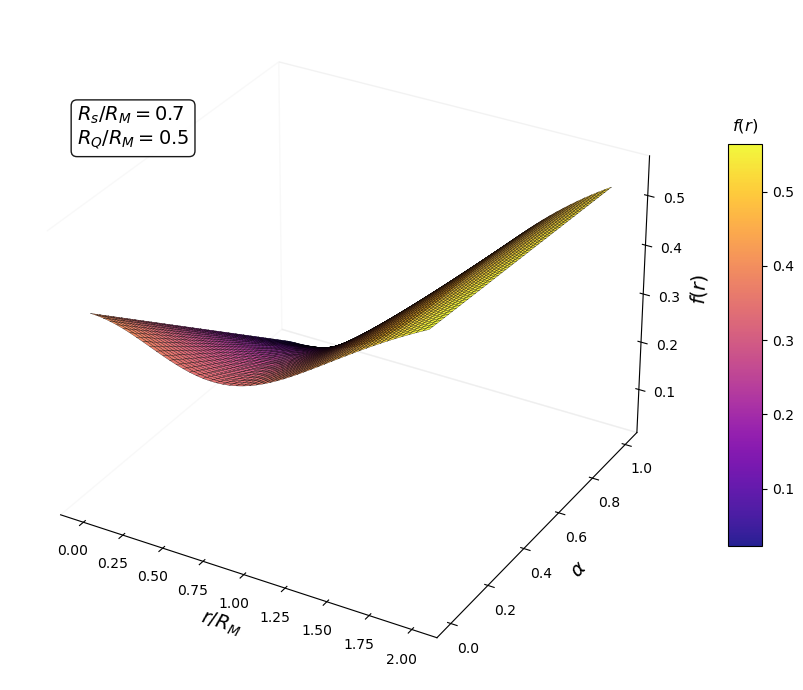}
        \label{fig:a0.7}
    }
    
    \caption{\small {Metric function $f(r, \alpha)$ for different values of the ratios $R_{\rm s}/R_{\rm M}$ and $R_{\rm Q}/R_{\rm M}$.}}
    \label{fig:f(r, alpha)}
\end{figure}
{\color{black}Fig. \ref{fig:f(r, alpha)} shows the three-dimensional profile of the the coherent quantum GD hairy metric function  $f(r,\alpha)$ as a function of the radial coordinate and the hairy parameter, for different choices of the ratios $R_{\rm s}/R_{\rm M}$ and $R_{\rm Q}/R_{\rm M}$. For the parameter sets shown in panels (a) and (b), the metric function exhibits a pronounced non-monotonic radial profile, characterized by a sharp decrease in the inner core region followed by a smooth increase towards an asymptotic value far from the core. The panels (c) and (d), as the parameters vary, reveal a qualitative transition to smoother and predominantly monotonic profiles for the metric function, as well as lesser values of $f(r, \alpha)$ in comparison with the other two panels. In all four panels, the metric function exhibits a milder dependence on the hairy parameter $\alpha$, indicating that the near-core structure of the solution is primarily controlled by the combined effect of $R_{\rm s}$ and $R_{\rm Q}$.}

{\color{black}For completeness, let us regard the coherent quantum GD hairy metric \eqref {eq:metric_ansatz}, whose components (\ref{mf}) 
are associated with the potential (\ref{mf0}) as 
\begin{align}\label{vgdh}
    V_\textsc{gdh}^{q}(r)
    =&
    -\frac{R_{\rm M}}{2r} \,\operatorname{erf}\left(\frac{r}{R_{\rm s}}\right) + \frac{R_{\rm Q}^2}{r \, R_{\rm s}} \, \operatorname{F}\left(\frac{r}{R_{\rm s}}\right) - 
    \frac{\alpha G_{\textsc{n} } \ell}{2r} \, \operatorname{erf} \left(\frac{r}{R_{\rm s}}\right) \nonumber\\[8pt]
    &+ \frac{\alpha R_{\rm M}}{8r} e^{R_{\rm s}^2/R_{\rm M}^2} \left[e^{2r/R_{\rm M}} \, \operatorname{erfc}\left(\frac{R_{\rm s}}{R_{\rm M}} + \frac{r}{R_{\rm s}}\right) - e^{-2r/R_{\rm M}} \, \operatorname{erfc}\left(\frac{R_{\rm s}}{R_{\rm M}} - \frac{r}{R_{\rm s}}\right)\right] \, . 
\end{align}}
Since the event horizon  $\mathring{r}$ is determined by the equation $f(\mathring{r})=0$, the coherent quantum GD hairy black hole metric (\ref{mf}) makes it correspond to 
\begin{equation}
    V_\textsc{gdh}^{q}(\mathring{r}_\pm)=-\frac{1}{2},
\end{equation}
whose solutions are not algebraic. Nevertheless, the qualitative structure of horizons can be inferred from the behavior of the potential. First,   $\displaystyle\lim_{r\to\infty}\erf(r/R_{\rm s})=1$ and the exponentially suppressed erfc terms vanish in this limit, yielding 
\begin{equation}
    \lim_{r\to\infty}V_\textsc{gdh}^{q}(r)=0.
\end{equation}
Hence, the coherent quantum GD hairy spacetime remains asymptotically flat. Depending on the combination of parameters $(R_{\rm M}, R_{\rm Q}, R_{\rm s}, \alpha)$, three scenarios arise. The first one comprises two distinct horizons (non-extremal), the second possibility regards a degenerate horizon (extremal or multiplicity-two solution), and the third case lacks an event horizon, in which case the singularity is naked in the classical sense, although resolved in the quantum-mechanical setup.

{\color{black}Near the origin, the Gaussian regularization and the quantum hair smooth the classical singularity, with 
\begin{equation}
    \lim_{r\to0^+}V_\textsc{gdh}^{q}(r)
    = \frac{R_{\rm Q}^2}{R_{\rm s}^2}
    - \frac{R_{\rm M}}{\sqrt{\pi}R_{\rm s}}-
    \frac{\alpha}{2} \left[\frac{R_{\rm M}}{\sqrt{\pi} R_{\rm s}} \left(\frac{2 G_{\textsc{n}}\ell}{R_{\rm M}}+1\right)
    - e^{R_{\rm s}^2/R_{\rm M}^2}\erfc\!\left(\frac{R_{\rm s}}{R_{\rm M}}\right)\right].\label{rto0}
\end{equation}}
Physically, this limit quantifies the strength of the quantum core and the hair near the origin. The classical ${r^{-1}}$ divergence is replaced by a finite value determined by $R_{\rm s}$ and $\alpha$, reflecting the resolution of the central singularity. If this central value reaches $-1/2$, a horizon emerges at or near the core. Otherwise, the quantum GD  hairy geometry may be horizonless, representing a compact quantum object where the would-be singularity is softened. The competition between the mass, charge, quantum core size, and hair amplitude $\alpha$ thus dictates the causal structure of the coherent quantum GD hairy spacetime, specifying how quantum effects and the additional GD source alter horizon formation and the inner geometry relative to classical RN black holes.

\section{Effective source in the coherent quantum GD hairy solution}
\label{sec:effective_source}

The quantum-corrected GD hairy black hole metric (\ref{mf}) can be written as
\begin{equation}
    f(r) = f^{q}(r) + f^{H}(r),
    \label{eq: quantum + hairy metric}
\end{equation}
where
\begin{equation}\label{qp}
    f^{q}(r) =
    1
    - \frac{R_{\rm M}}{r}\,
    \erf\!\left(\frac{r}{R_{\rm s}}\right)
    + \frac{R_{\rm Q}^{2}}{r^{2}}\,
    \frac{2r}{R_{\rm s}}\,
    \F\!\left(\frac{r}{R_{\rm s}}\right),
\end{equation}
represents the Gaussian-regularized quantum correction to the
RN geometry \cite{Antonelli:2025mcv}, {\color{black}while the GD hairy contribution reads
\begin{align}\label{hp}
f^{H}(r)
&=
\frac{\alpha R_{\rm M} e^{R_{\rm s}^2/R_{\rm M}^2}}{4r}
\Bigg[
e^{2r/R_{\rm M}}
\erfc\!\left(\frac{R_{\rm s}}{R_{\rm M}} + \frac{r}{R_{\rm s}}\right)
-
e^{-2r/R_{\rm M}}
\erfc\!\left(\frac{R_{\rm s}}{R_{\rm M}} - \frac{r}{R_{\rm s}}\right)
\Bigg]
\nonumber\\
&\quad
- \frac{\alpha G_\textsc{n}\ell}{r}
\erf\!\left(\frac{r}{R_{\rm s}}\right).
\end{align}}
The coherent quantum GD hairy metric function \eqref{eq: quantum + hairy metric} does not correspond to any known vacuum solution of the Einstein field equations. Nevertheless, the presence of the quantum core scale $R_{\rm s}$ suggests that the coherent quantum GD hairy geometry
is sourced by an effective matter distribution describing a
quantum-corrected extended object rather than a point singularity.
It is therefore natural to reconstruct the stress-energy tensor
directly from the Einstein field equations,
\begin{equation}\label{tmunu1}
    T^\mu_{\ \nu}
    =
    \frac{1}{8\pi G_\textsc{n}}
    \left(R^\mu_{\ \nu}-\frac12 Rg^\mu_{\ \nu}\right)
    =
    \mathrm{diag}(-\rho, p_r, p_t, p_t),
\end{equation}
in terms of the energy density, the radial pressure, and the tangential pressure associated with the coherent quantum GD hairy black hole metric
(\ref{eq: quantum + hairy metric}). {\color{black} Using the lower bound condition over the hairy parameter $\ell = M/e^2$, defined in Eq. \eqref{eq: lower bounds over l}, these quantities read
\begin{align}
    \rho(r) &= -p_r(r)
    = \frac{1 - f(r) - r f'(r)}{8 \pi G_\textsc{n} r^2}\nonumber\\[8pt]
    &= \rho_{\rm M} \,\frac{R_{\rm s}^2}{r^2} \, e^{-r^2/R_{\rm s}^2} \left[1 + \frac{\alpha}{2} \left(1 + \frac{1}{e^2}\right)\right] - 2 \rho_Q(r) \frac{r^2}{R_{\rm s}^2} \left(1 - \frac{2r}{R_{\rm s}} \operatorname{F}\left(\frac{r}{R_{\rm s}}\right)\right) \nonumber \\[8pt]
    &\;\;\;\;\; - \frac{\alpha \, e^{R_{\rm s}^2/R_{\rm M}^2}}{16 \pi \, G_{\textsc{n}} \, r^2} \left[e^{2r/R_{\rm M}} \operatorname{erfc}\left(\frac{R_{\rm s}}{R_{\rm M} } + \frac{r}{R_{\rm s}}\right) + e^{-2r/R_{\rm M}} \operatorname{erfc}\left(\frac{R_{\rm s}}{R_{\rm M} } - \frac{r}{R_{\rm s}}\right)\right],
    \label{eq: total energy density}
\end{align}}
where
\begin{equation}\label{pqr}
    \rho_M \equiv \frac{M}{2 \pi^{3/2} R_{\rm s}^3},
    \qquad
    \rho_Q(r) \equiv \frac{Q^2}{8 \pi r^4}.
\end{equation} 
\textcolor{black}{The relation $\rho=-p_r$ implies that the effective source reconstructed from Einstein's equations possesses a radial equation-of-state parameter
\begin{equation}
\omega_r \equiv \frac{p_r}{\rho}=-1.
\end{equation}
This vacuum-like behavior is a characteristic feature of several regular black hole geometries and is often associated with the emergence of an effective de Sitter core at short distances. In the present framework, however, it does not arise from the introduction of an additional matter sector. Instead, it follows directly from the coherent-state description of the gravitational field and from the Gaussian smearing encoded by the scale $R_{\rm s}$. The suppression of high-momentum modes softens the  UV  concentration of energy that would otherwise generate the classical central singularity, replacing it with an extended quantum core. 
The corresponding negative radial pressure provides an effective repulsive contribution in the deep interior, counterbalancing the gravitational attraction that drives the classical collapse toward divergent curvature. As a result, the central region behaves as an anisotropic quantum vacuum whose properties are determined by the coherent-state structure of the geometry. This interpretation is further supported by the fact that, when the regularity condition \eqref{eq30} is satisfied, both the curvature invariants and the effective energy density remain finite at the origin. Therefore, the regular core may be viewed as a self-consistent manifestation of the underlying quantum gravitational condensate encoded in the coherent-state formalism rather than as the consequence of an externally prescribed exotic matter source.}

The first term on the right-hand side of Eq. (\ref{eq: total energy density}) corresponds to a Gaussian mass distribution of width
$R_{\rm s}$, the second one reproduces the regularized electromagnetic
contribution, and the remaining terms encode the nontrivial
hairy corrections. The Gaussian factor ensures that the mass density
is exponentially suppressed at large distances and finite at the
origin, while the charge contribution is softened relative to the
classical $r^{-4}$ behavior due to the Dawson function term.
The hairy parameter $\alpha$ controls the strength of deviations from
the pure coherent quantum RN core.

{\color{black}Besides, the tangential pressure takes the form
\begin{align}
    p_t(r) &=
    \frac{2 f'(r) + r f''(r)}{16 \pi G_\textsc{n} r}\nonumber\\[8pt]
    &=
    \rho_{\rm M} \, e^{-r^2/R_{\rm s}^2} \left[1 + \frac{\alpha}{2} \left(1 + \frac{1}{e^2}\right)\right] - 2 \rho_{\rm Q}(r)\frac{r^5}{R_{\rm s}^5} \left[\frac{R_{\rm s}}{r} - \left(2 - \frac{R_{\rm s}^2}{r^2}\right) \operatorname{F}\left(\frac{r}{R_{\rm s}}\right)\right] \nonumber \\[8pt]
    &+ \frac{\alpha \, e^{R_{\rm s}^2/R_{\rm M}^2}}{16 \pi \, G_{\textsc{n}}\, R_{\rm M} \, r} \left[e^{2r/R_{\rm M}} \operatorname{erfc}\left(\frac{R_{\rm s}}{R_{\rm M} } + \frac{r}{R_{\rm s}}\right) - e^{-2r/R_{\rm M}} \operatorname{erfc}\left(\frac{R_{\rm s}}{R_{\rm M} } - \frac{r}{R_{\rm s}}\right)\right].
    \label{eq: total tangential pressure}
\end{align}}
\!Conversely to the classical RN case, all components of the
effective stress-energy tensor \eqref{tmunu1} remain finite as $r \to 0$.
In particular, the tangential pressure  \eqref{eq: total tangential pressure} approaches a constant
value at the origin, evincing the presence of a regular
anisotropic fluid core rather than a curvature singularity.
The difference between $p_r$ and $p_t$
in Eqs. \eqref{eq: total energy density} and \eqref{eq: total tangential pressure} encodes the intrinsic anisotropy induced by the combined
quantum smearing and hair contributions.
The quantum-corrected GD hairy solution  may therefore be interpreted as sourced by
a self-gravitating anisotropic quantum fluid whose characteristic
length scale is set by $R_{\rm s}$ and whose deviations from the quantum RN case are governed by the hairy parameter $\alpha$.

\begin{figure}[H]
    \includegraphics[scale=0.45]{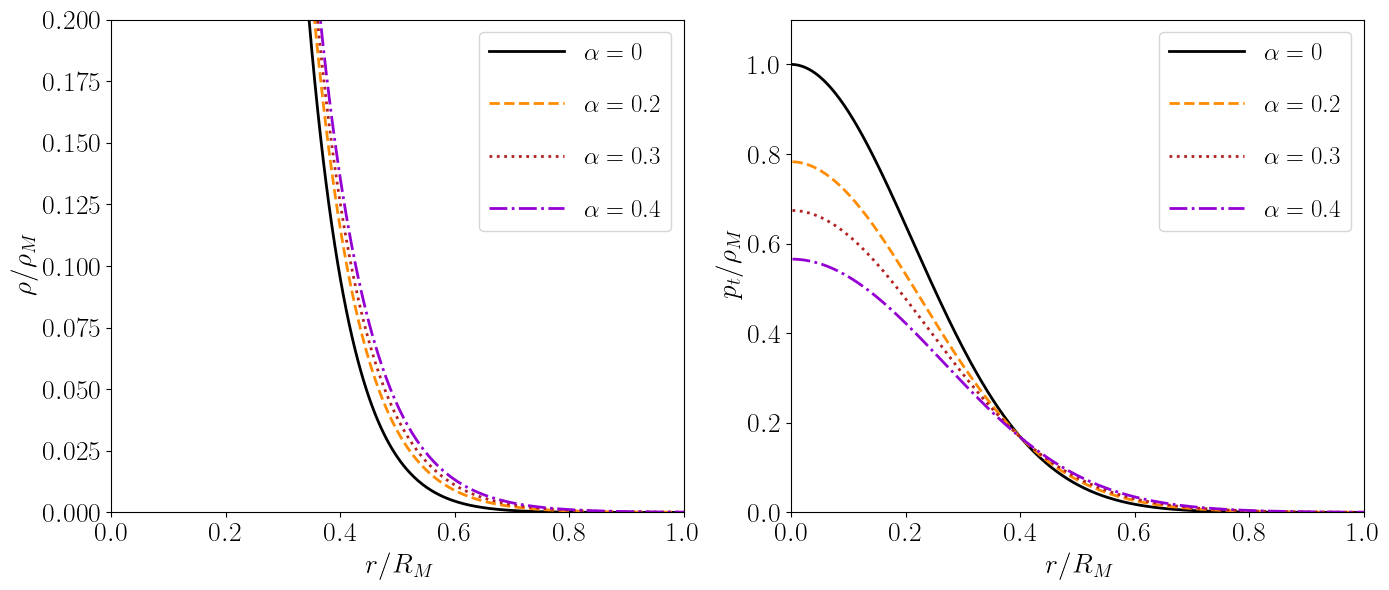}
    \caption{\small {\color{black} Normalized energy density (left) and tangential pressure (right) with respect to $r/R_{\rm M}$, for $R_{\rm s}/R_{\rm M} = 0.3$ and $R_{\rm Q}/R_{\rm M} = 0.1$, and using the minimal value of $\ell$.}}
    \label{fig: density and tangential pressure plots}
\end{figure}
{\color{black}Fig. \ref{fig: density and tangential pressure plots} displays the normalized effective energy density $\rho/\rho_M$ (left panel) and the normalized tangential pressure $p_t/\rho_M$ (right panel) as functions of $r/R_{\rm M}$ for different values of the hairy parameter $\alpha$ and an arbitrary pair of $R_{\rm s}/R_{\rm M}$ and $R_{\rm Q}/R_{\rm M}$. As $\alpha$ increases, both quantities are suppressed in the inner region, while their radial profiles remain monotonic and rapidly decay toward zero at large radii. This behavior indicates that increasing $\alpha$ weakens the effective matter content near the quantum core of the coherent quantum GD hairy distribution, without altering the asymptotic structure of the configuration.}

We can now scrutinize the behavior of the total energy density and tangential pressure in Eqs. \eqref{eq: total energy density} and \eqref{eq: total tangential pressure},   respectively, for two important regimes. The first one regards  $r/R_{\rm s} \gg 1$, corresponding to a large distance from the quantum core, whereas the second relevant regime consists of $r \to 0^+$, as close as possible to the center.  
For the coherent quantum GD hairy black hole, the quantum part of both the energy density and tangential pressure functions, $\rho^q(r)$ and $p^q_t(r)$,  respectively, are known to asymptotically approach $\rho_{\rm Q}(r)$ for $r/R_{\rm s}\gg 1$, emulating the results obtained for the quantum RN part for the whole coherent quantum GD hairy black hole  \cite{Antonelli:2025mcv}.   In addition,  since the Dawson function can be expanded as 
\begin{equation}
    \lim_{{r}/{R_{\rm s}} \gg 1}\operatorname{F}\left(\frac{r}{R_{\rm s}}\right) = \frac{{R_{\rm s}}}{2r} + \frac{{R^3_{\rm s}}}{4r^3} + \cdots,
\end{equation}
 the quantum energy density can be reduced to
\begin{eqnarray}
    \rho^q(r) \approx 2 \rho_Q(r) \frac{r^2}{R_{\rm s}^2}\left[1 - 2 \frac{r}{R_{\rm s}} \left(\frac{R_{\rm s}}{2r} + \frac{R_{\rm s}^3}{4r^3}\right)\right] =  \rho_Q(r),
\end{eqnarray}
and using the fact that the complementary error function has the following asymptotic behavior $\displaystyle\lim_{x\to\infty}\operatorname{erfc}(x)=0$, whereas $\displaystyle\lim_{x\to -\infty}\operatorname{erfc}(x)=2$, 
{\color{black}the total energy density in the limit of $r/R_{\rm s} \gg 1$ becomes 
\begin{equation}
    \rho(r) = \rho_{\rm Q}(r) - \frac{\alpha \, e^{R_{\rm s}^2/R_{\rm M}^2}}{8 \pi G_{\textsc{n}}} \; \frac{e^{-2r/R_{\rm M}}}{r^2}. 
    \label{eq: density function far from the quantum core}
\end{equation}
Similarly, the total tangential pressure at large distances from the quantum core is given by
\begin{equation}
    p_t(r) = \rho_{\rm Q}(r) - \frac{\alpha \, e^{R_{\rm s}^2/R_{\rm M}^2}}{8 \pi G_{\rm N} R_{\rm M}}\, \frac{e^{-2r/R_{\rm M}}}{r}. \label{eq: tangential pressure far from the quantum core}
\end{equation}}
Eq. \eqref{eq: density function far from the quantum core} shows that the energy density decays exponentially and with a $r^{-2}$ rate, while still carrying the hair charge $\alpha$ effects for long distances. The tangential pressure in Eq. \eqref{eq: tangential pressure far from the quantum core} presents a Yukawa-like decaying behavior, while also carrying the hair charge $\alpha$ along. {\color{black}If we substitute the explicit form of $\rho_{\rm Q}(r)$ defined in Eq. \eqref{pqr} in the expressions for the energy density and tangential pressure, the energy--momentum tensor components for the hairy black hole solution outside the event horizon are recovered \cite{Ovalle:2020kpd}. This emphasizes that the coherent quantum hairy black hole model is consistent with the classical geometry, which is recovered far from the quantum core.}

In the limit for $r \to 0^+$, the quantum energy density and tangential pressure terms are reduced to \cite{Antonelli:2025mcv}
\begin{eqnarray}
    \rho^q(r) = -p^q_r(r) &=& \left[\rho_{\rm M} - 2\rho_{\rm Q}(R_{\rm s})\right]\frac{R_{\rm s}^2}{r^2} + \mathcal{O}(1), \label{eq: density function near zero for the quantum part}\\[8pt]
    p_t^q(r) &=& \rho_{\rm M} - 4 \rho_{\rm Q}(R_{\rm s}) + \mathcal{O}(r^2).
\end{eqnarray}
{\color{black}The expansion of the hairy term of the total energy density \eqref{eq: total energy density} around $r = 0$ yields
\begin{equation}
    \lim_{r\to 0^+} \rho^H(r) = \left[-\frac{e^{R_{\rm s}^2/R_{\rm M}^2}}{4 \pi G_{\rm N} R_{\rm s}^2} \, \operatorname{erfc}\left(\frac{R_{\rm s}}{R_{\rm M}}\right) + \rho_{\rm M} \left(1 + \frac{1}{e^2}\right) \right] \frac{\alpha}{2} \, \frac{R_{\rm s}^2}{r^2} + \mathcal{O}(1) \label{eq: hairy part of the density function near zero}.
\end{equation}}
{\color{black}The total energy density when $r \to 0^+$ corresponds to the combination of Eqs. \eqref{eq: density function near zero for the quantum part} and \eqref{eq: hairy part of the density function near zero}, so that it becomes
\begin{equation}
\lim_{r \to 0^+} \rho(r) \simeq \left[\rho_{\rm M}\left(1 + \frac{\alpha}{2}\left(1 + \frac{1}{e^2}\right)\right) - 2 \rho_{\rm Q}(R_{\rm s}) - \frac{\alpha \, e^{R_{\rm s}^2/R_{\rm M}^2}}{8 \pi G_{\rm N} \, R_{\rm s}^2} \, \operatorname{erfc}\left(\frac{R_{\rm s}}{R_{\rm M}}\right)\right] \frac{R_{\rm s}^2}{r^2}\, .\label{eq: density function near zero}
\end{equation}}

{\color{black}Similarly, the expansion of the hairy term of the tangential pressure is given by 
\begin{equation}
    \lim_{r \to 0^+} p_t^{H}(r) = \frac{\alpha}{2} \rho_{\rm M} \left[\left(1 + \frac{1}{e^2}\right) + 2\frac{R_{\rm s}^2}{R_{\rm M}^2} \left(\frac{\sqrt{\pi} R_{\rm s}}{R_{\rm M}} \operatorname{erfc}\left(\frac{R_{\rm s}}{R_{\rm M}}\right) e^{R_{\rm s}^2/R_{\rm M}^2} - 1\right)\right]
\end{equation}}
Grouping it with the quantum part, {\color{black}the total tangential pressure can be written, up to $\mathcal{O}(r^2)$, as
\begin{align}
\lim_{r \to 0^+} p_t(r) &\simeq \rho_{\rm M} \left[1 + \frac{\alpha}{2}\left(1 + \frac{1}{e^2}\right)\right] 
    - 4\rho_{\rm Q}(R_{\rm s})+ \nonumber\\[8pt] 
    & \qquad \qquad \qquad \qquad + \alpha\,\rho_{\rm M}\,\frac{R_{\rm s}^2}{R_{\rm M}^2} \left[\frac{\sqrt{\pi} R_{\rm s}}{R_{\rm M}} \operatorname{erfc}\left(\frac{R_{\rm s}}{R_{\rm M}}\right) e^{R_{\rm s}^2/R_{\rm M}^2} - 1\right].\label{eq: tangential pressure function near zero}
\end{align}}
{\color{black}Eq. \eqref{eq: density function near zero} shows that the energy density and radial pressure diverge at $r = 0$, unless the coefficients cancel $r^{-2}$. Even though these components still carry a singularity at the origin, it is a much milder one compared to the classical RN geometry, since the volume integrals are always finite due to the being locally integrable for $r > 0$.} Meanwhile, the tangential pressure remains finite, due to Eq. \eqref{eq: tangential pressure function near zero}, which can also be realized in Fig. \ref{fig: density and tangential pressure plots}.

From the total energy density \eqref{eq: total energy density}, it is possible to compute the Misner--Sharp mass for the quantum hairy geometry, reading 
\begin{equation}
    m(r) = 4\pi \int_0^r \rho(\mathsf{r}) \; \mathsf{r}^2 \; d\mathsf{r}. 
\end{equation}
Integration performed over the quantum energy density $\rho^q(x)$ yields the result found in Ref. \cite{Antonelli:2025mcv}, which can be written as
\begin{equation}
    m^q(r) = M \operatorname{erf} \left(\frac{r}{R_{\rm s}}\right) - \frac{Q^2}{R_{\rm s}} \operatorname{F} \left(\frac{r}{R_{\rm s}}\right).
\end{equation}
{\color{black}Integration over the hairy energy density $\rho^H(r)$ yields the hairy contribution to the total Misner--Sharp mass, which can be read as
\begin{eqnarray}\label{msm}
    \!\!\!\!\!\!\!\! m^H(r) &\!=\!&  -\frac{\alpha}{4}  M e^{R_{\rm s}^2/R_{\rm M}^2}\left[e^{2r/R_{\rm M}} \operatorname{erfc}\left(\frac{r}{R_{\rm s}} \!+\!\frac{R_{\rm s}}{R_{\rm M}}\right) \!+\! e^{-2r/R_{\rm M}} \operatorname{erfc}\left(\frac{r}{R_{\rm s}}\!-\!\frac{R_{\rm s}}{R_{\rm M}}\right) \!-\!2 e^{-2r/R_{\rm M}}\right] \nonumber \\
    && +\frac{\alpha M}{2e^2} \operatorname{erf}\left(\frac{r}{R_{\rm s}}\right),
\end{eqnarray}
where we used the lower bound over the hairy parameter $\ell = M/e^2$ \eqref{eq: lower bounds over l}.

The Misner--Sharp mass for the quantum-corrected GD hairy solution can be written as 
\begin{eqnarray}
\!\!\!\!\!\!\!\!\!\!\!  m(r) &\!=\!&  M\left(1 + \frac{\alpha}{2e^2}\right) \operatorname{erf}\left(\frac{r}{R_{\rm s}}\right) - \frac{Q^2}{R_{\rm s}} \operatorname{F} \left(\frac{r}{R_{\rm s}}\right) \nonumber\\
&&    -\frac{\alpha}{4}  M e^{R_{\rm s}^2/R_{\rm M}^2}\!\left[e^{2r/R_{\rm M}} \operatorname{erfc}\left(\frac{r}{R_{\rm s}} \!+\!\frac{R_{\rm s}}{R_{\rm M}}\right) \!+\! e^{-2r/R_{\rm M}} \operatorname{erfc}\left(\frac{r}{R_{\rm s}}-\frac{R_{\rm s}}{R_{\rm M}}\right) \!-\!2 e^{-2r/R_{\rm M}}\right].  \label{eq: misner-sharp mass}
\end{eqnarray}}
For large values of $r$, the quantum corrected part $m^q(r)$ yields the ADM mass of the system \cite{Antonelli:2025mcv}, namely
\begin{equation}
    \lim_{r \to + \infty} m^q(r) = M. 
\end{equation}
{\color{black}However, the hairy correction terms do not vanish in the  $r \to + \infty$ limit, yielding a more intricate expression for large distances. In fact, the Misner--Sharp mass for the quantum hairy corrected geometry yields 
\begin{equation}
    \lim_{r \to + \infty}m(r) = M  \left(1 + \frac{\alpha}{2e^2}\right), \label{eq: asymptotic mass}
\end{equation}
which is consistent with observing the coherent quantum  GD hairy black hole very far from the quantum core, resembling the classical GD hairy limit \cite{Ovalle:2020kpd}. 
For a tiny characteristic size of the quantum core, the Misner--Sharp mass reduces to
\begin{equation}\label{msm1}
    \lim_{R_{\rm s} \to 0^+} m(r) = M  \left(1 + \frac{\alpha}{2e^2}\right) - \frac{Q^2}{2r} + \frac{\alpha M}{2} e^{-r/G_\textsc{n} M}.
\end{equation}
}

\begin{figure}[H]
    \subfigure[$\;\;R_{\rm Q}/R_{\rm M} = 0.3.$]{
        \includegraphics[width=0.48\textwidth]{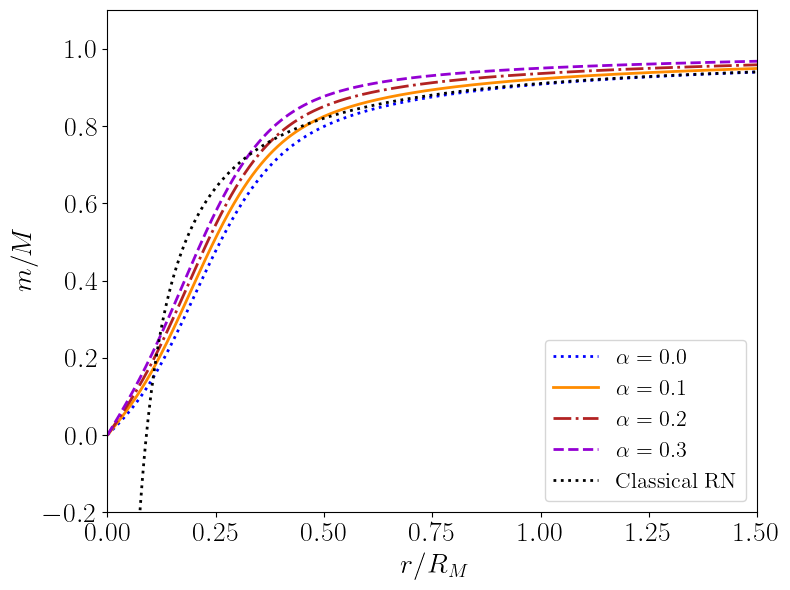}
        \label{fig:mx_b01}}
    \hfill
    \subfigure[$\;\;R_{\rm Q}/R_{\rm M} = 0.4.$]{
        \includegraphics[width=0.48\textwidth]{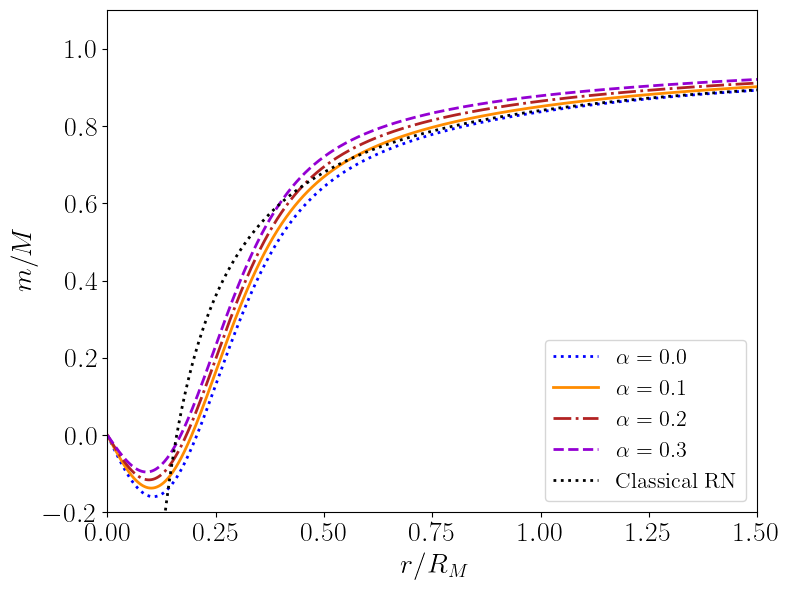}
        \label{fig:mx_b03}}
    \caption{\small {Behavior of the normalized mass function $m(r)/M$ for fixed $R_{\rm s}/R_{\rm M}=0.2$ and $R_{\rm Q}/R_{\rm M} = 0.3$ (left) and $R_{\rm Q}/R_{\rm M}=0.4$ (right).}}
    \label{fig:mass function plots}
\end{figure}
{\color{black}Fig. \ref{fig:mass function plots} shows the radial behavior of the normalized mass function $m(r)/M$ for different values of the hairy parameter $\alpha$, with fixed $R_{\rm s}/R_{\rm M}=0.2$ and two representative values of $R_{\rm Q}/R_{\rm M}$. For $R_{\rm Q}/R_{\rm M}=0.3$ (left panel), the normalized mass function $m(r)/M$ increases monotonically from the origin and sharply approaches its asymptotic value, with larger $\alpha$ resulting in larger total masses due to the hair charge. On the other hand, for $R_{\rm Q}/R_{\rm M}=0.4$ (right panel), the mass function develops a shallow negative region at small radii, whose depth decreases as a function of $\alpha$, while all curves converge to an asymptotic mass at large radial distance, whose value depends on the hair charge $\alpha$, as stated in Eq. \eqref{eq: asymptotic mass}. In both plots, we verify the finiteness of the mass function $m(r)/M$ of the coherent quantum hairy black hole, different of  the divergent mass function of the classical RN black hole.}

\section{Resolving the singularity at the origin}
\label{sec:regularity}
The metric function for the quantum-corrected GD hairy geometry $f(r) = 1 + 2 V^q_\textsc{gdh}(r)$ is analytic. 
Therefore, it is possible to expand it around $r = 0$ through a power series as proposed in Ref. \cite{Antonelli:2025mcv}:
\begin{equation}
    f(r) = \sum_{n = 0}^{\infty} f_n r^n \label{eq: power series of the metric},
\end{equation}
where
$\displaystyle
    f_n = \frac{f^{(n)}(0)}{n!}.
$ To check whether the quantum-corrected GD hairy geometry is regular around the origin, it is necessary to analyze the curvature invariants, which can be conveniently written in terms of the expansion coefficients of Eq. \eqref{eq: power series of the metric}. These can be read up to $\mathcal{O}(r)$, respectively, as \cite{Antonelli:2025mcv}
\begin{subequations}
\begin{eqnarray}\label{kr-1}
    \!\!\!\!\!\!\!\!\!\!\!\!\lim_{r\to0^+}R &=& -\frac{2}{r^2}(f_0 - 1) - \frac{6}{r} f_1 - 12f_2,\\
\!\!\!\!\!\!\!\!\!\!\!\!\!\!\!\!\lim_{r\to0^+}R_{\phantom{\mu}\nu}^{\mu} R_{\phantom{\mu}\mu}^\nu &=& \frac{2}{r^4}(f_0 - 1)^2 + \frac{8}{r^3} f_1 (f_0 - 1) + \frac{2}{r^2}\left[6f_2(f_0 - 1) + 5f_1^2\right] \nonumber \\
    &&+\frac{4}{r}\left[4f_3(f_0-1) + 9f_1f_2\right] + 4\left[5 f_4(f_0 - 1) + 14 f_1 f_3 + 9f_2^2\right],\\
\!\!\!\!\!\!\!\!\!\!\!\!\lim_{r\to0^+}R_{\mu\nu\rho\sigma}R^{\mu\nu\rho\sigma} &=& \frac{4}{r^4}(f_0 - 1)^2 + \frac{8}{r^3} f_1 (f_0 - 1) + \frac{8}{r^2} \left[f_2(f_0 - 1) + f_1^2\right]\nonumber \\
    &&+\frac{8}{r}\left[f_3(f_0-1) + 3 f_1f_2\right] + 8\left[f_4(f_0 - 1) + 4 f_1 f_3 + 3f_2^2\right].\label{kr1}
\end{eqnarray}
\end{subequations}
When $f_0 = 1$ and $f_1 = 0$, the three quantities (\ref{kr-1}) -- (\ref{kr1}) are finite: 
\begin{equation}
    R = -12f_2, \qquad \qquad R_{\phantom{\mu}\nu}^{\mu} R_{\phantom{\mu}\mu}^\nu = 36 f_2^2, \qquad \qquad R_{\mu\nu\rho\sigma}R^{\mu\nu\rho\sigma} = 24 f_2^2.
\end{equation}
To check the geometry regularity, it is sufficient to calculate the $f_0$, $f_1$, and $f_2$ coefficients. Specializing to the quantum-corrected GD hairy metric function, given by $f(r)$ in Eq. \eqref{eq: quantum + hairy metric}, it follows that
\begin{subequations}
\begin{eqnarray}\label{f01}
   \!\!\! \!\!\!\!\!\!\!\!\!\!\!f_0 &=& 1 + \frac{2}{R_{\rm s}}\left(\frac{R_{\rm Q}^2}{R_{\rm s}} - \frac{R_{\rm M}}{\sqrt{\pi}}\right) - \frac{2 \alpha G_\textsc{n}\ell}{\sqrt{\pi} R_{\rm s}} - \alpha\left[\frac{R_{\rm M}}{\sqrt{\pi} R_{\rm s}}-\operatorname{erfc}\left(\frac{R_{\rm s}}{R_{\rm M}}\right) e^{R_{\rm s}^2/R_{\rm M}^2}\right], \\
   \!\!\! \!\!\!\!\!\!\!\!\!\!\!f_1 &=& 0, \label{eq: f0} \\
   \!\!\!\!\!\! \!\!\!\!\!\!\!\!f_2 \!&\!=\!&\! -\frac{4}{3R_{\rm s}^3}\left(\frac{R_{\rm Q}^2}{R_{\rm s}} \!-\! \frac{R_{\rm M}}{2 \sqrt{\pi}}\right) \!+\! \frac{2\alpha G_\textsc{n} \ell}{3 \sqrt{\pi} R_{\rm s}^3} \!-\! \frac{\alpha}{3} \left[\frac{2}{\sqrt{\pi} R_{\rm M} R_{\rm s}} \!-\! \frac{R_{\rm M}}{\sqrt{\pi} R_{\rm s}^3} \!-\! \frac{2}{ R_{\rm M}^2} \operatorname{erfc}\left(\frac{R_{\rm s}}{R_{\rm M}}\right) e^{R_{\rm s}^2/R_{\rm M}^2}\right].
\end{eqnarray}
\end{subequations}
Hence, two possible scenarios can be read off, depending on whether $f_0 = 1$ or $f_0 \neq 1$. If the first condition is satisfied, a regular configuration at $r = 0$ sets in, corresponding to a single, doubly-degenerate horizon. 

{\color{black}Replacing the condition $f_0 = 1$ in Eq. \eqref{f01} and manipulating it, while using the extremal condition for the hair charge $\ell$ given in Eq. \eqref{eq: lower bounds over l}, we end up with the following equation:
\begin{equation}
    \frac{R_{\rm Q}^2}{R_{\rm M}^2} = \frac{R_{\rm s}}{\sqrt{\pi} R_{\rm M}} \left[1 + \frac{\alpha}{2} \left(1 + \frac{1}{e^2}\right)\right] - \frac{\alpha}{2} \frac{R_{\rm s}^2}{R_{\rm M}^2} e^{R_{\rm s}^2/R_{\rm M}^2} \operatorname{erfc} \left(\frac{R_{\rm s}}{R_{\rm M}}\right). \label{eq30}
\end{equation}}
\!\textcolor{black}{It is worth emphasizing that the regularity condition \eqref{eq30} plays a dual role in the present construction. Besides leading finite curvature invariants, it also removes the leading divergence in the effective energy density (\ref{eq: density function near zero}). {\color{black}This can be explicitly seen by manipulating the coefficient in Eq. \eqref{eq: density function near zero} and using the explicit forms of $\rho_{\rm M}$ and $\rho_{\rm Q}(R_{\rm s})$, leading to
\begin{eqnarray}
    \frac{R_{\rm M}^2}{R_{\rm s}^4} \frac{1}{4\pi G_{\textsc{n}}} \left[\frac{R_{\rm s}}{\sqrt{\pi} R_{\rm M}} \left(1 + \frac{\alpha}{2}\left(1 + \frac{1}{e^2}\right)\right) - \frac{R_{\rm Q}^2}{R_{\rm M}^2} - \frac{\alpha}{2} \frac{R_{\rm s}^2}{R_{\rm M}^2} e^{R_{\rm s}^2/R_{\rm M}^2}\operatorname{erfc} \left(\frac{R_{\rm s}}{R_{\rm M}}\right)\right] \label{eq: coefficient of the energy density}\, .
\end{eqnarray}
By a simple inspection, one clearly realizes that substituting the expression \eqref{eq30} in the $R_{\rm Q}^2/R_{\rm M}^2$ term in Eq. \eqref{eq: coefficient of the energy density} makes the coefficient null. As a consequence, this leads to a null and finite value of the energy density $\rho(r)$ at $r=0$, according to Eq. \eqref{eq: density function near zero}. The fact that Eq. \eqref{eq30} determines the regularization of both the energy density function $\rho(r)$ and the curvature invariants is not merely an algebraic coincidence. This condition was originally obtained from the purely geometric requirement that the curvature invariants remain finite at the origin, without any additional constraints on the matter source. It determines precisely the combination of parameters that leads to the vanishing of the curvature singularity. Remarkably, the same relation automatically removes the leading divergence in the reconstructed energy density.} Therefore, the regularity of the geometry and that of the effective matter sector emerge simultaneously from the same condition, {\color{black} being a direct consequence of the regularity of the spacetime geometry}. This behavior is consistent with recent developments on regular black holes, where the singularity resolution has been achieved through different mechanisms, including phantom scalar fields, nonlinear DBI matter sectors, and effective loop-quantum-gravity corrections \cite{Gonzalez:2025yjm,Parvez:2025wtq,Belfaqih:2024vfk}. Although the microscopic origin of the regularization differs among these approaches, they state that a physically meaningful nonsingular spacetime requires not only finite curvature invariants but also a well-behaved effective source in the high-curvature regime. In our framework, this requirement is automatically satisfied as long as Eq.~(\ref{eq30}) holds. Conversely, away from this condition, the effective energy density has singular behavior near the origin, signaling the breakdown of the coherent-state interpretation of the regularized core.}

\textcolor{black}{In addition, it is worth emphasizing that Eq.~(\ref{eq30}) plays a central role in the regularization mechanism. This relation among the characteristic scales of the solution ensures the simultaneous cancellation of the leading short-distance singularities in both the curvature invariants and the effective energy density reconstructed from Einstein's equations. Conversely, departures from Eq.~(\ref{eq30}) generally spoil these cancellations and reintroduce the leading singular contribution in the near-origin expansion of the effective source. Therefore, the regularity of the solution is tied to the fulfillment of Eq.~(\ref{eq30}), which identifies a distinguished regular subset of the model parameter space. Whether this relation remains preserved under quantum fluctuations or higher-order corrections is a question that depends on the microscopic theory underlying the coherent-state description and lies beyond the scope of the present effective analysis.}

{\color{black}The fact that the regularity condition also regularizes the source represents a nontrivial confirmation of the consistency of Einstein's equations in the effective quantum regime in the high curvature region near the core. Since the energy-momentum tensor $T^{\mu}_{\phantom{\mu} \nu}$ is reconstructed from the geometry through the Einstein tensor $G^{\mu}_{\phantom{\mu}\nu}$ in Section \ref{sec:effective_source}, we should expect a confluence between the geometry and matter terms in the model. The fact that the same algebraic condition over the set of parameters $(R_{\rm Q}, R_{\rm s}, R_{\rm M}, \alpha$) simultaneously eliminates divergences near the origin in both sectors is a relevant  nontrivial result that ensures the validity of Einstein's field equations.}

{\color{black}The limit $R_{\rm s}/R_{\rm M} \ll 1$ yields 
\begin{equation}
    \frac{R_{\rm s}}{R_{\rm M}} = \frac{2\sqrt{\pi} \left(R_{\rm Q}/R_{\rm M}\right)^2}{2 + \alpha\left(1 + 1/e^2\right)}. \label{eq: regularity for small Rs}
\end{equation}}
Therefore, for $R_{\rm s}/R_{\rm M} \ll 1$, it is possible to find values of $R_{\rm s}$ that make the geometry of  the quantum GD hairy black hole to be regular at $r = 0$, for certain values of $R_{\rm Q}$, $R_{\rm M}$ and $\alpha$.

\begin{figure}
    \centering
    \includegraphics[scale=0.5]{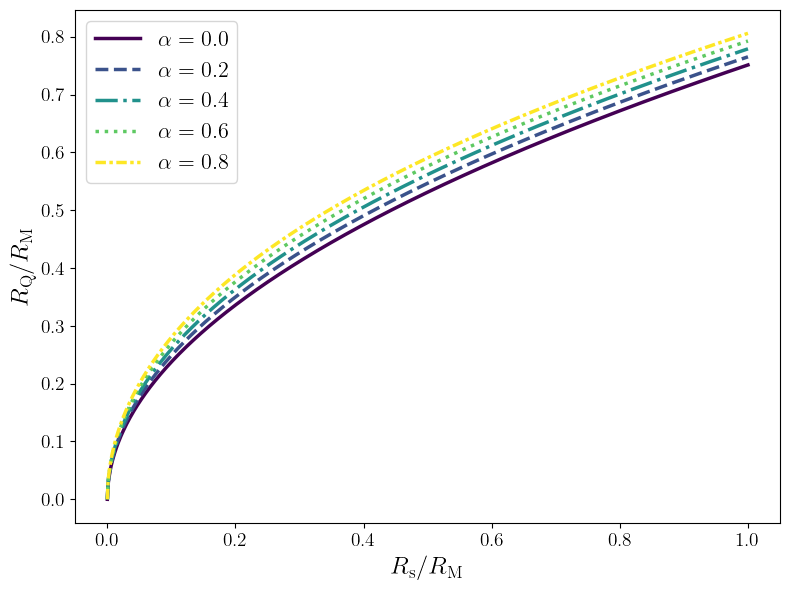}
    \caption{\small {Values of $R_{\rm Q}/R_{\rm M}$ in terms of $R_{\rm s}/R_{\rm M}$ that provide a regular geometry at the origin.}}
    \label{fig50}
\end{figure}

In a more general approach, we can solve Eq. \eqref{eq30} numerically for fixed values of $\alpha$, and find a set of solutions for $R_{\rm Q}/R_{\rm M}$ in terms of $R_{\rm s}/R_{\rm M}$. This can be visualized in Fig. \ref{fig50}.
Fig.~\ref{fig50} illustrates the relation between the normalized parameters $R_{\rm Q}/R_{\rm M}$ and $R_{\rm s}/R_{\rm M}$ for different values of the hairy parameter  $\alpha$. In all cases, $R_{\rm Q}/R_{\rm M}$ increases monotonically with $R_{\rm s}/R_{\rm M}$, defining a well-behaved parameter space for the solutions. Increasing $\alpha$ shifts the curves upward, indicating that stronger GD hairy effects enhance the allowed charge scale $R_{\rm Q}/R_{\rm M}$ for a fixed quantum core radius $R_{\rm s}/R_{\rm M}$.

On the other hand, if the case $f_0 \neq 1$ is considered, Eq. \eqref{eq30} does not hold anymore, and the quantum-corrected GD hairy geometry contains an integrable singularity at $r=0$. As pointed out in Ref. \cite{Antonelli:2025mcv}, this emulates  the case for the coherent quantum RN  geometry, and even though the curvature invariants diverge, the integrable singularity is weaker than in the classical RN geometry \cite{Lukash:2013}. 
The integrability of this singularity is due to the nature of the energy density function $\rho(r)$ \eqref{eq: total energy density} in the coherent quantum GD hairy geometry. Even though $\rho(r)$ is not regular at $r=0$, as can be seen clearly in Fig. \ref{fig: density and tangential pressure plots}, it yields a finite Misner--Sharp mass \eqref{eq: misner-sharp mass}, which can also be seen in Fig. \ref{fig:mass function plots}. The fact that the mass function vanishes at $r=0$ can be explained by the finiteness nature of the potential $V^q_{\textsc{gdh}}$ due to the quantum corrections of the coherent-state approach \cite{Arrechea:2025,Lukash:2013}.

\section{Strong-field photon dynamics and spectral signatures  of coherent quantum GD hairy black holes}

\label{geo1}
The coherent quantum GD hairy black hole geometry significantly modifies the RN geometry. As a consequence, the existence of a quantum core and corrections due to the hairy parameter in the strong-curvature regime produce observable effects, allowing observational tests to constrain the scale of modifications to GR.  

In this section, we study the dynamical properties of massless particles in the geometry associated with coherent quantum GD hairy black holes and investigate how the hairy parameter affects observational properties beyond the classical RN geometry. Ref.  \cite{Antonelli:2025mcv} investigated how a purely coherent quantum RN black hole geometry affects observational properties of photons. This geometry is a particular case of coherent quantum  GD hairy black holes for $\alpha=0$ and $\ell= 0$. Ref.  \cite{Urmanov:2024qai} considered a similar approach for a quantum Schwarzchild-like geometry and analyzed the dynamics of both massive and massless particles. 

The standard approach to investigating geodesics is followed, considering the motion of point-like particles on the equatorial plane (corresponding to the azimuthal angle $\theta = \pi/2$) without loss of generality, due to the rotational isometry. The two Killing vectors corresponding to two constants of motion for particles on geodesics are the particle energy and its angular momentum, both quantities measured per unit mass, respectively given by \cite{Boehmer:2009bmu}  
\begin{subequations}
\begin{gather}
     E = f(r) \dot t  \label{eq: geodesics energy}\, ,\\[8pt]
    L = r^2 \dot \phi \, . \label{eq: geodesics angular momentum}
\end{gather}
\end{subequations}
Then, the geodesic equation reads:
\begin{eqnarray}\label{scn}
    \dot r^2 &=& E^2 - V_\textsc{eff}(r),
    \end{eqnarray} where 
    \begin{eqnarray}\label{scn1}  V_\textsc{eff} &=& f(r) \left(\frac{L^2}{r^2} - \mu\right), 
\end{eqnarray}
with $\mu$ being the norm of the tangent vector to the geodesic, taking $\mu = -1, 0, 1$ whether the geodesic is timelike, null, or spacelike, respectively, and the dot sign denoting the derivative with respect to the affine parameter along the geodesic. Throughout this section, we will address the dynamics of photons in the spacetime defined by the coherent quantum GD hairy black hole metric, probing for observable astrophysical properties.

\begin{figure}[H]
    \centering
    \includegraphics[width=0.48\textwidth]{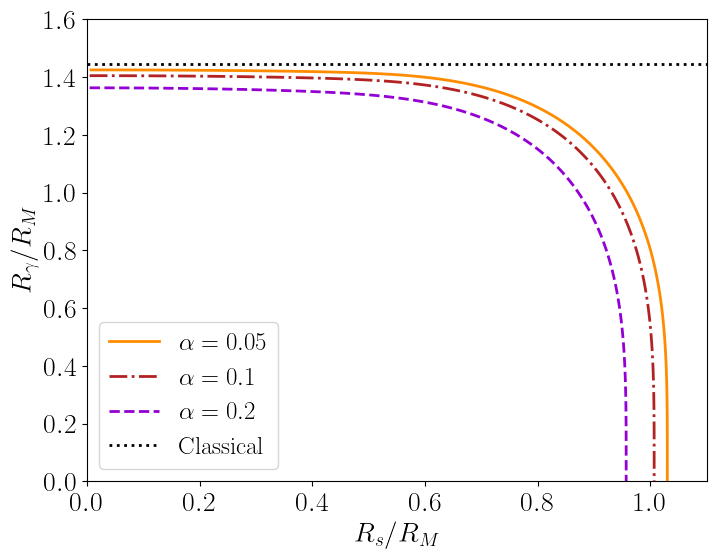}
    \hfill
    \includegraphics[width=0.48\textwidth]{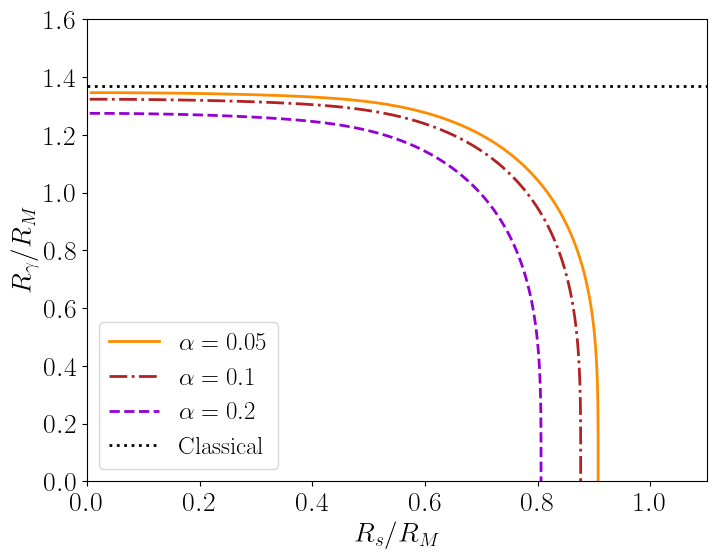}
    \caption{\small {Photon radius $R_\gamma$ in units of $R_{\rm M}$ as a function of $R_{\rm s}/R_{\rm M}$ for different values of the hairy parameter $\alpha$. In the left panel, we set $R_{\rm Q}/R_{\rm M} = 0.2$ and in the right panel, we set $R_{\rm Q}/R_{\rm M} = 0.3$. The classical value for the photon radius is plotted as a dashed black line.}}
    \label{fig40}
\end{figure}

Since photons are massless particles, the geodesics are null. In this case, one considers $\mu = 0$ in Eq.  (\ref{scn1}). 
A particularly relevant quantity to obtain is the photon ring radius $R_\gamma$. It can be derived through the condition $V'_\textsc{eff}(R_\gamma) = 0$, which implies
\begin{equation}
   R_\gamma \, f'(R_\gamma) - 2f(R_\gamma) = 0. \label{eq: photon ring condition}
\end{equation}
We can split Eq. \eqref{eq: photon ring condition} into two parts. The first one consists of the quantum contribution $f^q(r)$ (\ref{qp}) of $f(r)$  whose solution was found in Ref. \cite{Antonelli:2025mcv}, whereas the second one comprises the hairy part $f^H(r)$ (\ref{hp}). It implies that Eq. \eqref{eq: photon ring condition} can be written as 
\begin{equation}
    [R_\gamma \, f'^q(R_\gamma) - 2f^q(R_\gamma)] + [R_\gamma \, f'^H(R_\gamma) - 2f^H(R_\gamma)] = 0.
\end{equation}
 The quantum part yields the following expression for $R_\gamma$:
\begin{equation}
    1 - \frac{R_{\rm Q}^2}{R_{\rm s}^2} + \frac{R_{\rm M}}{\sqrt{\pi} R_{\rm s}} \, \exp \left(-\frac{R_\gamma^2}{R_{\rm s}^2}\right) + \frac{R_{\rm Q}^2 (3R_{\rm s}^2 + 2R_\gamma^2)}{R_\gamma R_{\rm s}^3} \, \operatorname{F} \left(\frac{R_\gamma}{R_{\rm s}}\right) - \frac{3 R_{\rm M}}{2 R_\gamma} \,\operatorname{erf}\left(\frac{R_\gamma}{R_{\rm s}}\right)=0. \label{eq: photon radius equation for the quantum part}
\end{equation}
{\color{black}The hairy component of the metric  \eqref{eq: quantum + hairy metric} yields the following terms containing the photon ring radius, where the lower bound condition for $\ell$ \eqref{eq: lower bounds over l} is already imposed: 
\begin{multline}
    \frac{3 \alpha R_{\rm M}}{2e^2 \, R_\gamma} \, \operatorname{erf} \left(\frac{R_\gamma}{R_{\rm s}}\right) -\frac{\alpha R_{\rm M}}{\sqrt{\pi} R_{\rm s}}\left(1 + \frac{1}{e^2}\right) \exp\left(-\frac{R_\gamma^2}{R_{\rm s}^2}\right) + \\[8pt]
    + \frac{\alpha}{2}\operatorname{erfc}\left(\frac{R_{\rm s}}{R_{\rm M}} + \frac{R_\gamma}{R_{\rm s}}\right) \exp\left(\frac{R_{\rm s}^2}{R_{\rm M}^2} + \frac{2R_{\gamma}}{R_{\rm M}}\right) \left(1- \frac{3R_{\rm M}}{2R_\gamma}\right) + \\[8pt]
    +\frac{\alpha}{2} \operatorname{erfc}\left(\frac{R_{\rm s}}{R_{\rm M}} - \frac{R_\gamma}{R_{\rm s}}\right) \exp\left(\frac{R_{\rm s}^2}{R_{\rm M}^2} - \frac{2R_{\gamma}}{R_{\rm M}}\right) \left(1+ \frac{3R_{\rm M}}{2R_\gamma}\right)=0. \label{eq: photon radius equation for the hairy part}
\end{multline}}
The photon ring equation for the coherent quantum GD hairy black hole  model is given by a sum of Eqs. \eqref{eq: photon radius equation for the quantum part} and \eqref{eq: photon radius equation for the hairy part}. It can only be solved through numerical methods, and the curves representing $R_{\gamma}/R_{\rm M}$ in terms of $R_{\rm s}/R_{\rm M}$ are displayed in Fig. \ref{fig40}.

{\color{black}Fig.~\ref{fig40} shows the photon ring radius $R_\gamma$ (in units of $R_{\rm M}$) as a function of $R_{\rm s}/R_{\rm M}$ for different values of the GD hairy parameter $\alpha$, with $R_{\rm Q}/R_{\rm M}=0.2$ (left panel) and $R_{\rm Q}/R_{\rm M}=0.3$ (right panel). For each value of $\alpha$, the photon ring becomes slightly smaller as the ratio $R_{\rm Q}/R_{\rm M}$ increases, while for fixed $R_{\rm Q}/R_{\rm M}$ the photon radius decreases with $\alpha$, indicating the photon ring decreases due to GD hair. At small and intermediate values of $R_{\rm s}/R_{\rm M}$, increasing $\alpha$ leads to photon radii smaller than the classical value and extends the admissible range of $R_{\rm s}/R_{\rm M}$ before the disappearance of circular photon orbits. However, for each fixed $\alpha$ there exists a critical regime in which the photon ring becomes smaller than that of the quantum RN black hole. In the weak-hair limit, all curves approach the classical photon radius, ensuring consistency with the standard case.}

Once the photon ring radius is determined, it is possible to find the critical impact parameter, defined as
\begin{equation}
    b_{\rm c} = \frac{R_{\gamma}}{\sqrt{f(R_{\gamma})}}. \label{eq: critical impact parameter bc}
\end{equation}

\begin{figure}[H]
    \centering
    \includegraphics[width=0.48\linewidth]{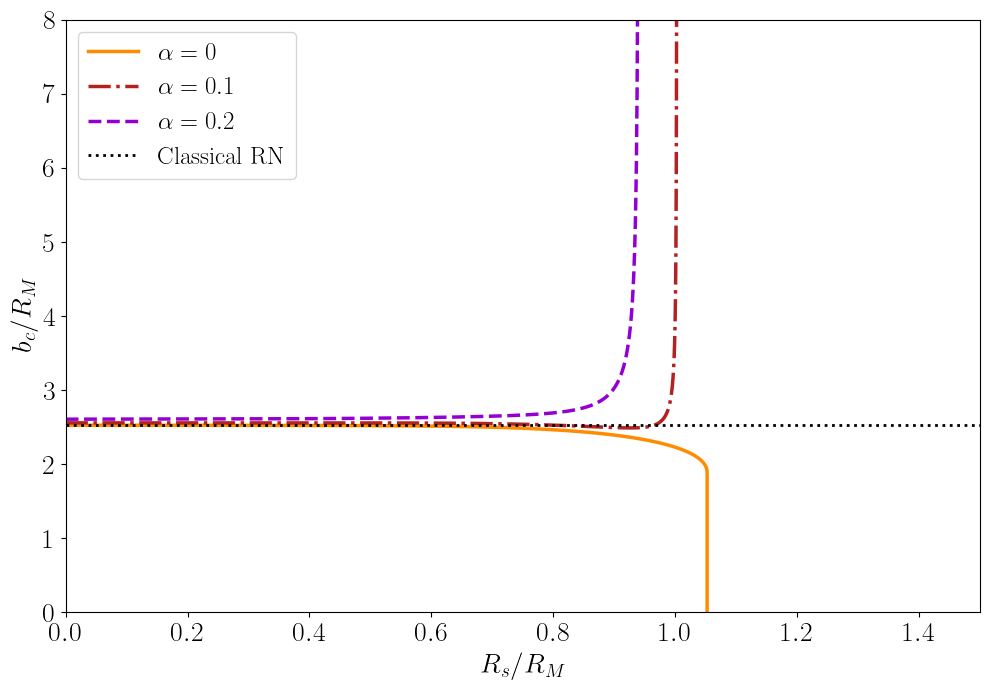}
    \hfill
    \includegraphics[width=0.48\linewidth]{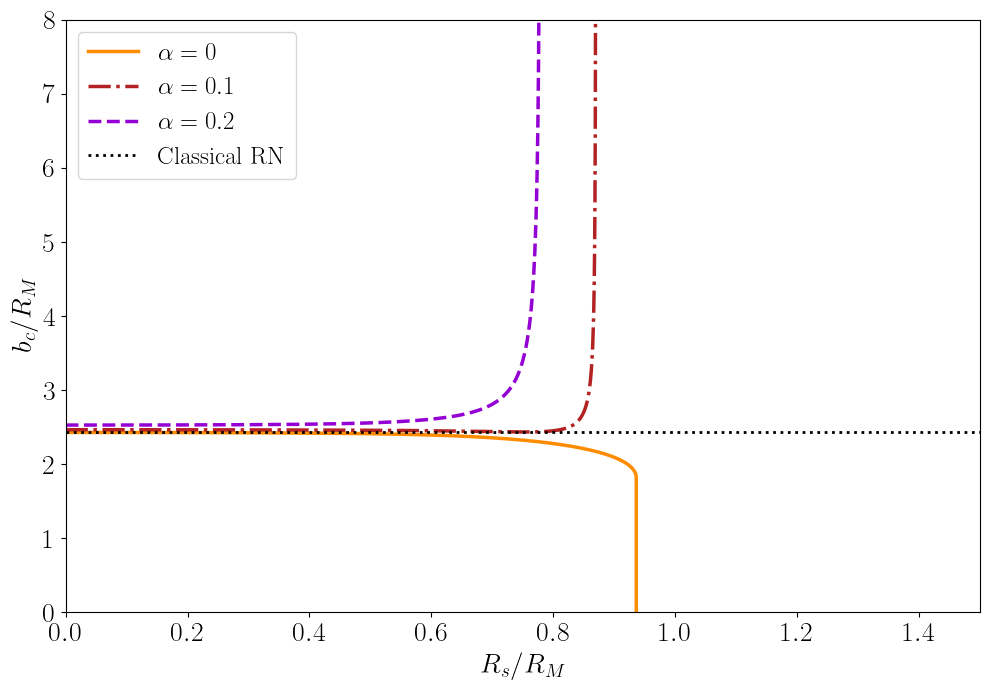}
    \caption{\small {Critical impact parameter $b_c$ in units of $R_{\rm M}$ with respect to $R_{\rm s}/R_{\rm M}$ for different values of the hairy parameter $\alpha$. In the left panel, we set $R_{\rm Q}/R_{\rm M} = 0.2$ and in the right panel, we set $R_{\rm Q}/R_{\rm M} = 0.3$. In both plots, the classical impact parameter value is shown.}}
    \label{fig: critical impact parameter}
\end{figure}

Eq. \eqref{eq: critical impact parameter bc} corresponds to the minimal impact parameter value $b = L/E$ below which the photon falls into the black hole. Since we have determined the photon radius $R_\gamma$ numerically, the critical impact parameter is obtained similarly. Fig. \ref{fig: critical impact parameter} shows the behavior of $b_c/R_{\rm M}$ with respect to $R_{\rm s}/R_{\rm M}$.

{\color{black} Fig. \ref{fig: critical impact parameter} displays the critical impact parameter $b_c$ (in units of $R_{\rm M}$) as a function of $R_{\rm s}/R_{\rm M}$ for different values of the hairy parameter $\alpha$, with $R_{\rm Q}/R_{\rm M}=0.2$ (left panel) and $R_{\rm Q}/R_{\rm M}=0.3$ (right panel). For comparison, the classical RN critical impact parameter and the hairless coherent quantum RN solution ($\alpha = 0$) are also shown. A clear qualitative distinction emerges between the cases with vanishing and non-vanishing GD hair. For $\alpha = 0$, $b_c$ is nearly indistinguishable from the classical RN value for lower values of $R_{\rm s}/R_{\rm M}$, and as the quantum core becomes larger, it starts to decrease, eventually vanishing at a critical value of $R_{\rm s}/R_{\rm M}$. This behavior indicates the existence of a maximum quantum-core size beyond which the spacetime no longer supports a non-vanishing critical impact parameter. For $\alpha \neq 0$, the GD hair modifies the critical impact parameter value even for lower $R_{\rm s}/R_{\rm M}$ configurations. Larger values of $\alpha$ increase the value of $b_c$ even when the quantum core is small, indicating that the presence of hair increases the minimum impact parameter required for photons to remain on critical orbits. Nevertheless, we still encounter a limit value of $R_{\rm s}/R_{\rm M}$ for which $b_c$ remains stable. Unlike the hairless case, where the critical impact parameter decreases and eventually vanishes, the presence of GD hair leads to a divergent behavior of $b_c$ as a critical value of $R_{\rm s}/R_{\rm M}$ is approached. Furthermore, this divergence occurs at progressively smaller values of $R_{\rm s}/R_{\rm M}$ as $\alpha$ increases, suggesting that the combined effects of the quantum core and GD hair reduce the range of admissible quantum-core sizes while simultaneously enhancing the critical impact parameter.}

\begin{figure}[H]
    \centering
    \includegraphics[width=0.48\linewidth]{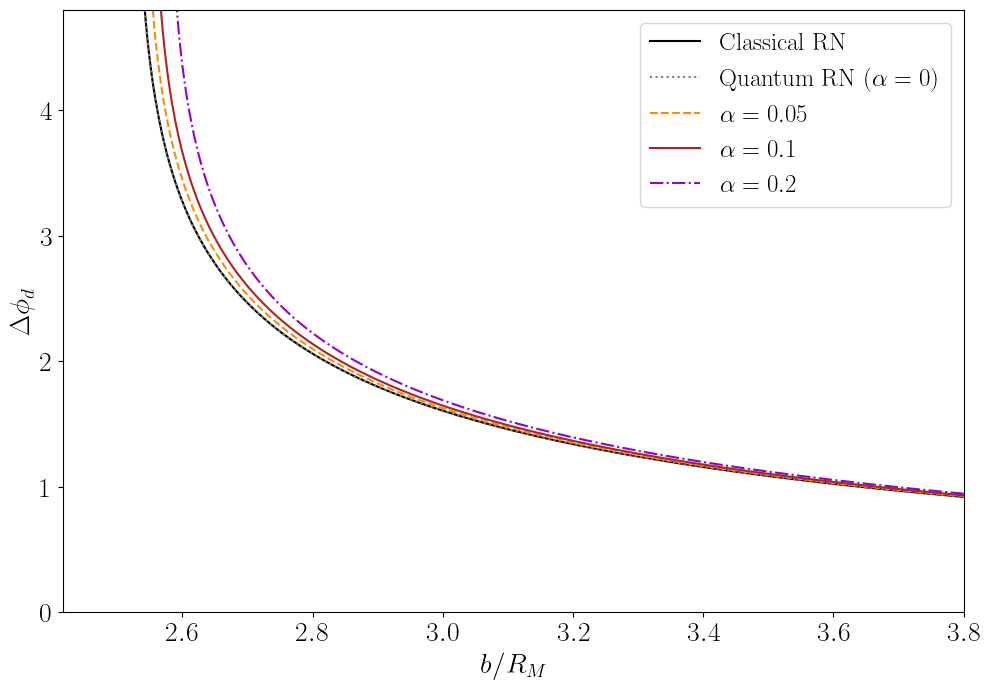}
    \hfill
    \includegraphics[width=0.48\linewidth]{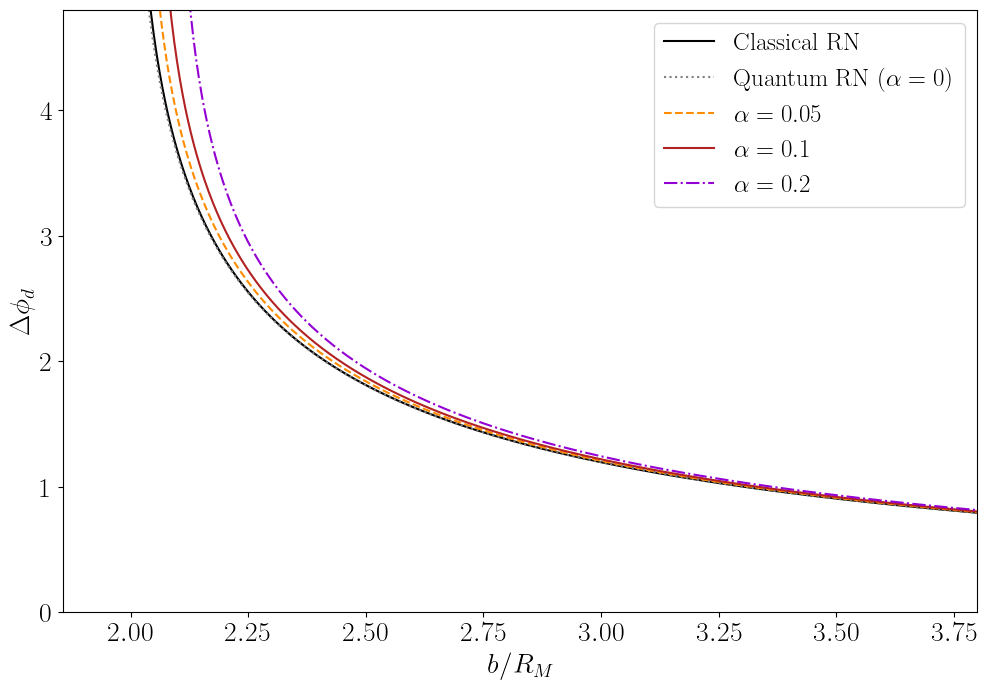}
    \caption{\small {Deflection angle $\Delta \phi_d = \Delta \phi - \pi$ of light due to  the coherent quantum GD hairy black hole, in terms of the normalized impact parameter. In both plots, we used $R_{\rm s}/R_{\rm M} = 0.1$. In the left panel, we set $R_{\rm Q}/R_{\rm M} = 0.2$, and in the right panel we set $R_{\rm Q}/R_{\rm M} = 0.5$.}} 
    \label{fig: light deflection plots}
\end{figure}

Another important observable quantity is the gravitational lensing due to the strong gravitational fields of black holes. Since the angular momentum is given by Eq. (\ref{eq: geodesics angular momentum}), the geodesics of photons can be used to quantify the angular deflection of light in the coherent quantum GD hairy black hole geometry, reading 
\begin{equation}
    \dot \phi = \pm \frac{\dot r}{r \sqrt{\displaystyle\frac{r^2}{b^2} - f(r)}}. \label{eq: ODE of the light deflection}
\end{equation} 
When $\dot r = 0$, the orbit presents a turning point at radius $r = r_0$. Imposing this position in Eq. \eqref{scn}, and using  Eq. \eqref{eq: geodesics angular momentum}, it is possible to determine that the impact parameter $b$ can be written in terms of the turning point $r_0$ as
\begin{equation}
    b = \frac{r_0}{\sqrt{f(r_0)}}. \label{eq: impact parameter in the turning point}
\end{equation}
To compute the angle variation of light $\Delta \phi$ due to the presence of the quantum-corrected GD hairy black hole, we must integrate Eq. \eqref{eq: ODE of the light deflection}, assuming that the light ray comes from infinity, reaches the turning point $r_0$, and drifts off toward infinity in space. This yields:
\begin{equation}
    \Delta \phi = 2 \int_{r_0}^{\infty} \frac{dr}{r \sqrt{\displaystyle \frac{r^2}{b^2} - f(r)}}. \label{eq: integral of the light deflection}
\end{equation}
The integral in Eq. \eqref{eq: integral of the light deflection} is computed numerically due to the high complexity of the metric function. The results are shown in Fig. \ref{fig: light deflection plots}, where the light deflection angle $\Delta \phi_d = \Delta \phi - \pi$ is plotted in terms of the normalized impact parameter $b/R_{\rm M}$, with the influence of the hairy parameter $\alpha$.

\noindent {\color{black} Fig. \ref{fig: light deflection plots} shows the light deflection angle as a function of the normalized impact parameter $b/R_{\rm M}$ for the coherent quantum GD hairy black hole, compared with the classical RN case, and the hairless coherent quantum RN solution. In both panels, the classical RN light deflection is almost indistinguishable from the deflection from the coherent quantum RN solution. Increasing the GD hair parameter $\alpha$ enhances the deflection angle at small impact parameters, primarily in the strong-deflection regime (small impact parameters), whereas all curves converge at large $b/R_{\rm M}$. The left and right panels indicate that a larger charge increases the deviation from the classical RN behavior in the strong-lensing regime.}

The observational properties studied here reveal that the GD hair charge yields relevant deviations from both the purely coherent quantum black hole model of \cite{Antonelli:2025mcv} and the classical RN black hole. This implies that astrophysical observations may be used to test the coherent quantum GD hairy black hole model.

\section{QNMs of coherent hairy black holes} \label{sec: quasinormal modes}

\textcolor{black}{QNMs provide one of the most powerful probes of black-hole spacetimes, since the ringdown spectrum is entirely determined by the underlying geometry \cite{KokkotasSchmidt:1999,Berti:2009kk}. As a consequence, even small deviations from the Schwarzschild or Kerr metrics can leave characteristic imprints on the oscillation frequencies and damping times, making QNMs a valuable tool for testing the coherent quantum GD hairy black hole, which combines two distinct sources of geometric modifications. The GD scalar hair, which alters the exterior gravitational field, and coherent quantum corrections, which regularize the near-horizon and interior regions. Since these two parameters affect different sectors of the quantum hairy GD geometry, it is not evident whether they produce independent, competing, or even complementary signatures in the QNM spectrum. The main goal of this section is therefore to investigate how their interplay modifies the effective scattering potential and the corresponding ringdown frequencies.}
 Ref. \cite{Cavalcanti:2022cga} investigated scalar perturbations of the classical GD hairy black hole and showed that the corresponding quasinormal frequencies differ from those of the Schwarzschild spacetime, demonstrating the influence of the GD hair on the effective scattering potential. In addition, Ref. \cite{Guimaraes:2025jsh} studied gravitational perturbations of the classical GD hairy solution and found deviations from the predictions of GR. More recently, Refs. \cite{Antonelli:2025mcv,Antonelli:2026qnm} analyzed QNMs of coherent quantum black-hole geometries and showed that quantum corrections also modify the ringdown spectrum. These previous results motivate investigating how the combined effects of GD hair and coherent quantum corrections influence the QNM spectrum.

Since the spacetime is electrically charged, gravitational and electromagnetic perturbations are coupled, making a complete analysis of spin-1 and spin-2 perturbations considerably more involved. Our goal here is instead to isolate the imprint of the quantum-corrected geometry itself. We therefore consider a minimally coupled, massless test scalar field propagating on the fixed background. Although such scalar perturbations are not directly observable GW  modes, they provide a clean and widely used probe of the effective geometry and allow us to quantify how the coherent quantum corrections and the GD hair modify the QNM spectrum. Since our main aim in this section is to show the unique signatures of coherent quantum hairy black holes compared to other geometries, we restrict ourselves to computing the QNMs of scalar perturbations. These can be treated as the QNMs of a minimally coupled, massless scalar field $\psi$ propagating in our fixed background. The field dynamics can be described by the Klein--Gordon equation,
\begin{eqnarray}
    \square \, \psi = \nabla^\mu \nabla_\mu \psi = 0\, . \label{eq: Klein--Gordon equation}
\end{eqnarray}
Solutions to Eq. \eqref{eq: Klein--Gordon equation} are given in terms of Fourier modes of frequency $\omega$ and spherical harmonics $Y^l_m(\theta, \phi)$ as \cite{Zhidenko:2005mv}
\begin{eqnarray}
    \psi_{\omega l m} (t,r,\theta, \phi) = \int d\omega\, \sum_{l,m} e^{-i\omega t} \frac{\Psi(r)}{r} Y^l_m (\theta, \phi)\, , 
\end{eqnarray}
where $l \in \mathbb{N}$ is the orbital number, $m \in \mathbb{Z}$ is the azimuthal number, with $- l \leq m \leq l$. Such decomposition is possible due to the spherical symmetry of our geometry, with the newly introduced function $\Psi(r)$ representing the radial term. At this point, it is convenient to introduce the usual generalized tortoise coordinates $r_\star$, associated with the region outside the event horizon, as
\begin{eqnarray}
    \frac{dr_\star}{dr} = f(r)^{-1}\, ,
\end{eqnarray}
with $f(r)$ in our case being the metric function defined in Eq. \eqref{eq: quantum + hairy metric}. Equation \eqref{eq: Klein--Gordon equation} can be recast as a Schrödinger-like equation, taking the form
\begin{eqnarray}
    \frac{d^2 \Psi(r_\star)}{dr_\star^2} + [\omega^2 - V_0 (r)] \Psi(r_\star) = 0\, , \label{eq: schroedinger like equation}
\end{eqnarray}
where $V_0(r)$ is an effective scalar potential given as
\begin{eqnarray}
    V_0(r) = f(r) \left[\frac{l(l+1)}{r^2} + \frac{f'(r)}{r}\right]\, . \label{eq: effective potential qnm}
\end{eqnarray}

In the case of the coherent quantum hairy black hole, the spacetime geometry depends on the set of parameters $(M,\alpha, Q, \ell, R_{\rm s})$, which are respectively the black hole mass, the three primary hair charges, and the size of the quantum core. Throughout this section, we shall employ geometric units and set $M = 1$. Besides, we also work with the saturated values for the hair charges $Q$ and $\ell$ (Eq. \eqref{eq: lower bounds over l}) to reduce the number of degrees of freedom of our system. This set of physical conditions reduces the spacetime parameters to the 2-tuple $(\alpha, R_{\rm s})$, and the metric function $f(r, \alpha, R_{\rm s})$ can be recast as
\begin{align}
    f(r,\alpha, R_{\rm s}) = 1 &- \frac{R_{\rm M}}{r} \left(1 + \frac{\alpha}{2e^2}\right) \operatorname{erf} \left(\frac{r}{R_{\rm s}}\right) + \frac{2 \alpha R_{\rm M}^2}{r \, R_{\rm s}\, e^2} \,\operatorname{F}\left(\frac{r}{R_{\rm s}}\right) + \nonumber\\[8pt]
    &+\frac{\alpha R_{\rm M}}{4r} \, e^{R_{\rm s}^2/R_{\rm M}^2} \left[e^{2r/R_{\rm M}} \operatorname{erfc}\left(\frac{R_{\rm s}}{R_{\rm M}} + \frac{r}{R_{\rm s}}\right) - e^{-2r/R_{\rm M}} \operatorname{erfc}\left(\frac{R_{\rm s}}{R_{\rm M}} - \frac{r}{R_{\rm s}}\right)\right]. \label{eq: metric function for qnm}
\end{align}

\begin{figure}
    \centering
    \includegraphics[width=0.6\linewidth]{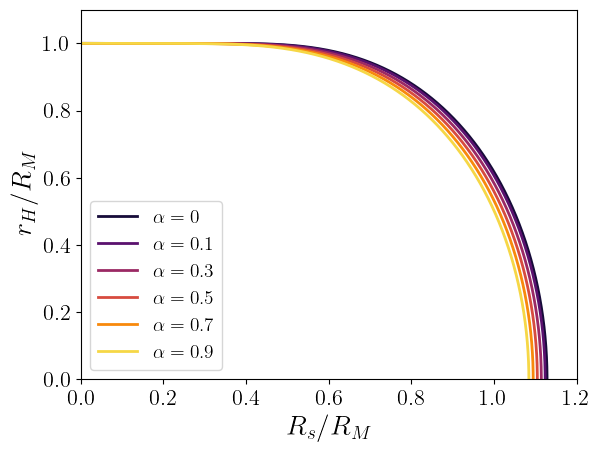}
    \caption{Event horizon position relative to the quantum core size of a coherent quantum hairy black hole, described by the metric \eqref{eq: metric function for qnm}, for different values of $\alpha$.}
    \label{fig:horizon_radius}
\end{figure}

The position of the black hole event horizon is an important feature when analyzing its QNMs, since the tortoise coordinate $r_\star$ maps the semi-infinite region from the horizon to infinity into the $(-\infty, +\infty)$ region \cite{KonoplyaZhidenko:2011}. When considering the spacetime described by the metric function \eqref{eq: metric function for qnm}, the event horizon can be found by computing  $f(r_H,\alpha, R_{\rm s}) = 0$, which has no analytical solution, although it can be solved numerically. Fig.  \ref{fig:horizon_radius} depicts the dependence of $r_H/R_{\rm M}$ with respect to $R_{\rm s}/R_{\rm M}$ for some values of $\alpha$. It is interesting to notice that for $0 < R_{\rm s}/R_{\rm M} \lesssim 0.4$ the event horizon remains practically unaltered relatively to the Schwarzschild radius. However, it is important to emphasize that even for low values of $R_{\rm s}/R_{\rm M}$ the event horizon cannot be considered precisely $r_H = 2$, due to the influence of the hair parameter $\alpha$, which slightly shifts the horizon radius upwards. The dependence of $\alpha$ becomes more noticeable with increasing values of the quantum core size, until it reaches a point where $R_{\rm s}$ surpasses the corresponding horizon. This suggests that such geometry could describe black hole mimickers under certain conditions, which is a feature found in other coherent quantum black hole models (see, e.g, Ref.  \cite{Antonelli:2026qnm}). \textcolor{black}{In particular, once the quantum core exceeds the event-horizon radius, no trapped region exists despite the object remaining compact and regular. Such configurations resemble horizonless compact objects, whose quasinormal spectra are expected to differ qualitatively from those of genuine black holes.}

\subsection{Effective potential}

\begin{figure}
    \centering
    \subfigure[\;$l=0$.]{
        \includegraphics[width=0.45\textwidth]{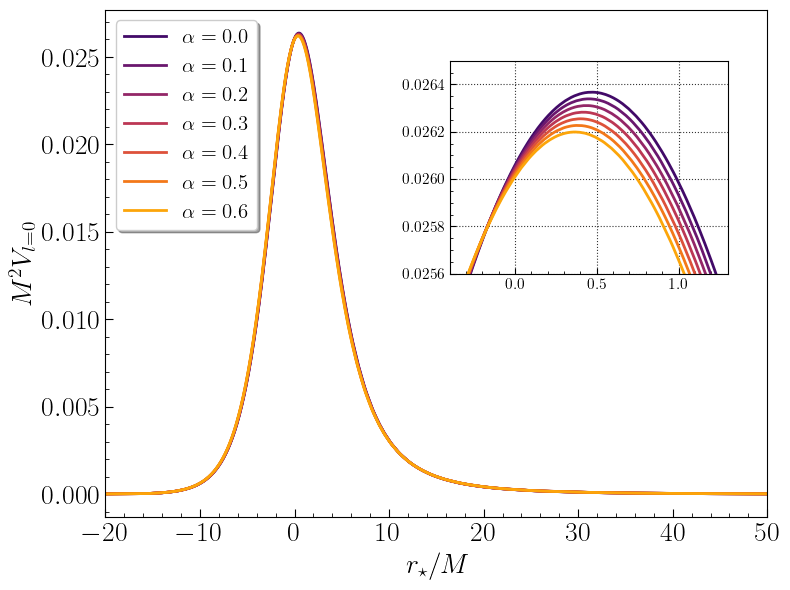}
        \label{fig:l=0}
    }
    \hfill
    \subfigure[\;$l=1$.]{
        \includegraphics[width=0.45\textwidth]{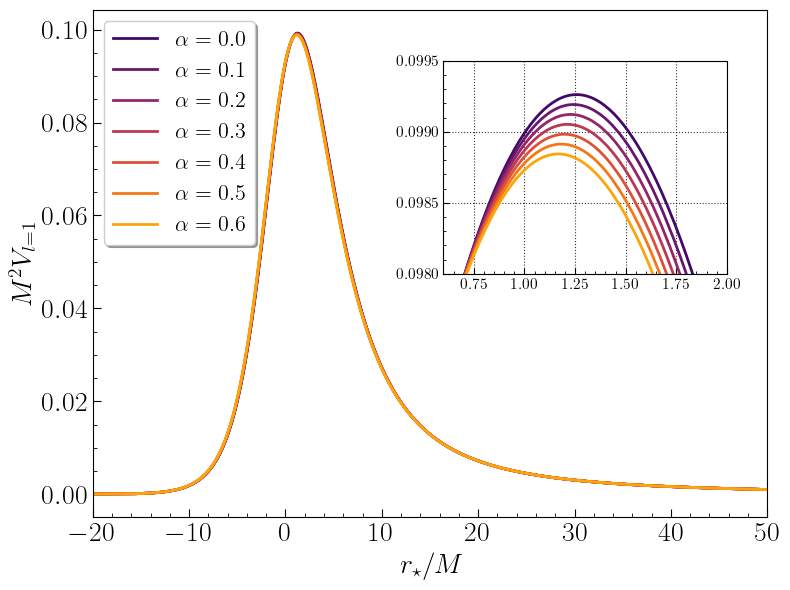}
        \label{fig:l=1}
    }
    
    \vspace{0.5cm} 
    
    \subfigure[\;$l = 2$.]{
        \includegraphics[width=0.45\textwidth]{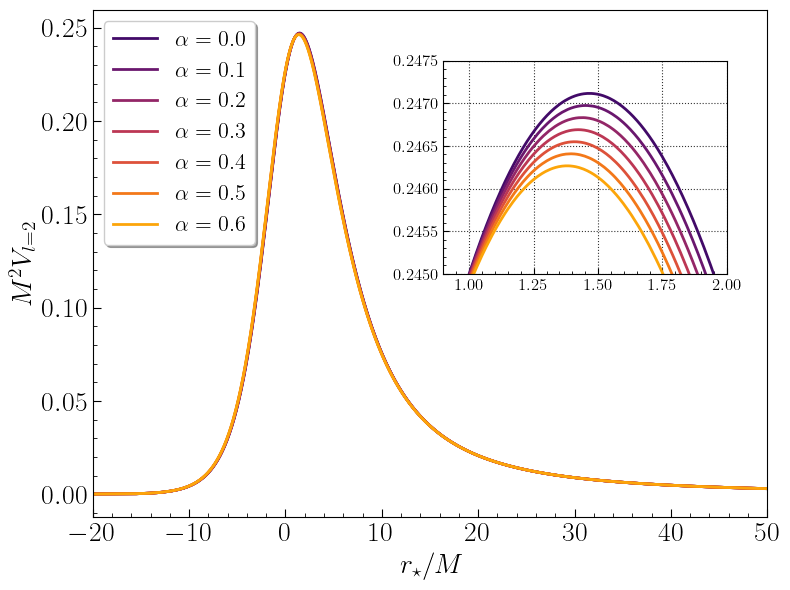}
        \label{fig:l=2}
    }
    \hfill
    \subfigure[\;$l = 3$.]{
        \includegraphics[width=0.45\textwidth]{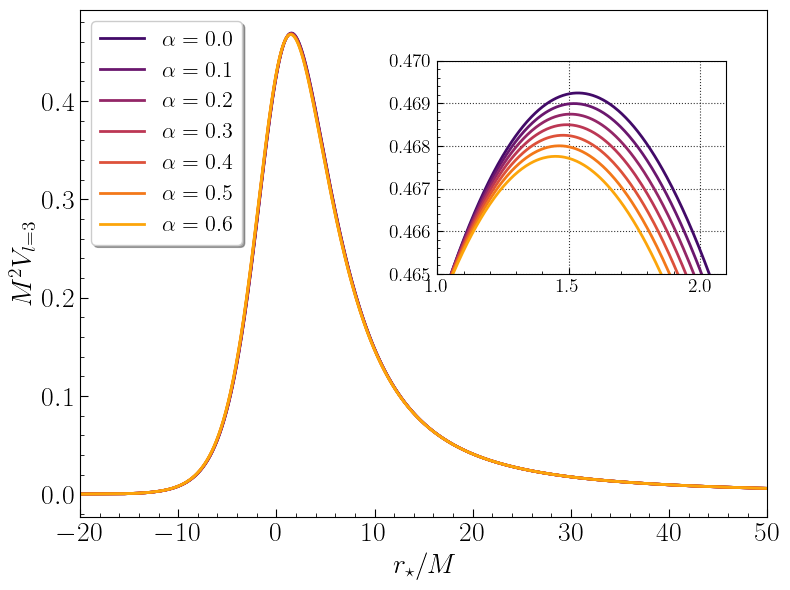}
        \label{fig:l=3}
    }
    
    \caption{\small {Effective potential for the metric \eqref{eq: metric function for qnm} in tortoise coordinates for $l=0,1,2,3$, and different values of the hair parameter $\alpha$. For every configuration, we set $M=1$, and used a fixed value of $R_{\rm s} = 0.2$.}}
    \label{fig: effective potential plot for qnm}
\end{figure}

Fig. \ref{fig: effective potential plot for qnm} depicts the dependence of $V_0(r)$ on the hair parameter $\alpha$, for a fixed quantum core size $R_{\rm s}=0.2$, and four different values of the orbital number $l$. \textcolor{black}{For $\alpha=0$ we recover the Schwarzschild effective potential. As the hair parameter increases, the potential barrier becomes progressively lower while its peak is displaced only slightly along the tortoise coordinate. This behavior is also observed in the classical GD hairy black hole \cite{Cavalcanti:2022cga}. However, the effect is considerably weaker in the coherent quantum geometry. From Fig.~\ref{fig: effective potential plot for qnm}, one finds that increasing the hair parameter from $\alpha=0$ to $\alpha=0.6$ lowers the potential maximum by an amount ranging from $0.01\%$ to $0.3\%$, depending on the value of $l$, whereas the position of the maximum changes only marginally. Thus, unlike the classical solution, where the GD hair substantially suppresses the scattering barrier, the coherent quantum corrections largely preserve its height while introducing only small quantitative modifications. Physically, increasing the GD hair parameter redistributes the effective gravitational field outside the horizon, weakening the curvature responsible for the trapping of scalar waves. As a consequence, the effective potential barrier becomes lower than in the Schwarzschild case.}

This attenuation effect over the black hole effective potential is also present in the classical GD hairy black hole geometry, as reported in Ref.  \cite{Cavalcanti:2022cga}, which  verified that for $\alpha = 0.6$ the potential peak is $\sim $ 10\% lower than the Schwarzschild case. However, in the current coherent quantum GD hairy black hole, such attenuation is not as sharp as it is in the classical picture. By inspecting Fig. \ref{fig: effective potential plot for qnm}, one can estimate that for $\alpha =0.6$ the largest potential peak is $\sim 0.3 \%$ lower than the peak for $\alpha = 0$, for $l=0$, implying that for objects within the coherent quantum black hole models the GD hair has a weaker influence over the effective potential.

\begin{figure}
    \centering
    \includegraphics[width=0.48\linewidth]{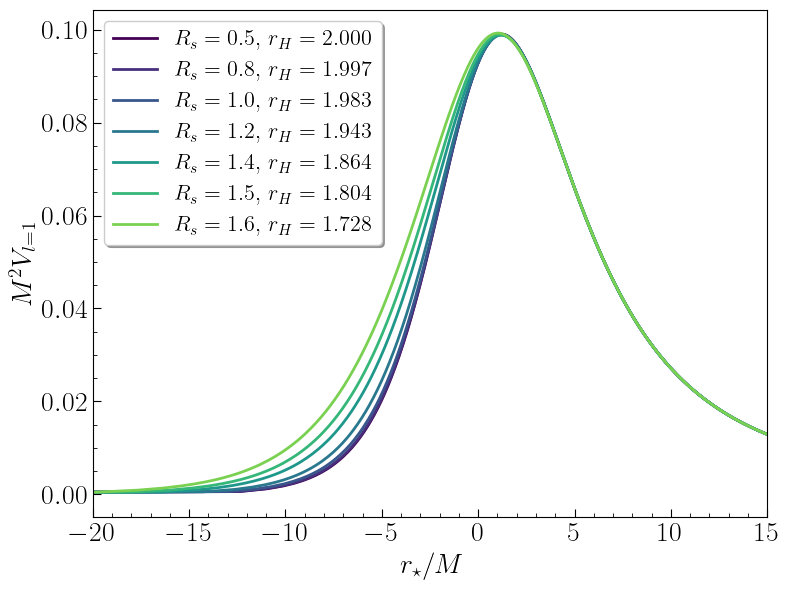}
    \hfill
    \includegraphics[width=0.48\linewidth]{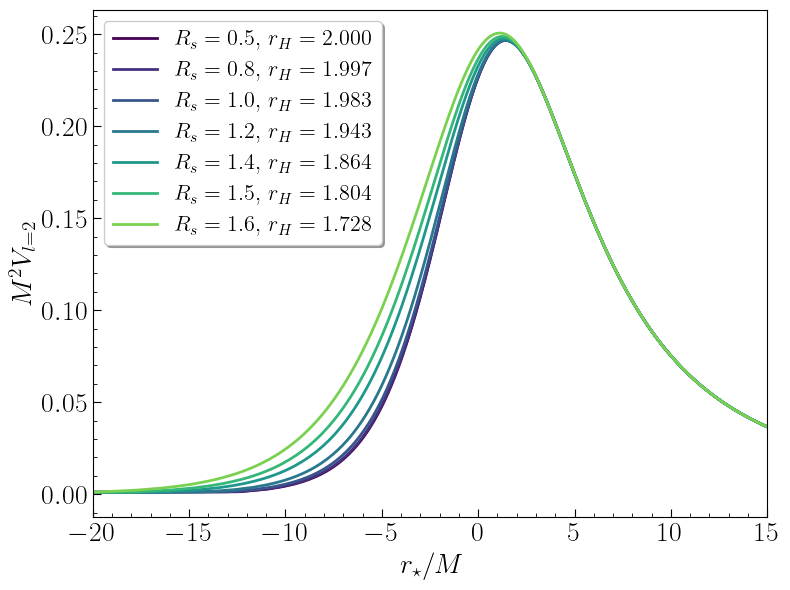}
    \caption{\small {Effective potential for the metric \eqref{eq: metric function for qnm} in tortoise coordinates for $l = 1$ (left) and $l = 2$ (right), and different values of quantum core size $R_{\rm s}$. For every configuration, we set $M=1$, and use a fixed value of $\alpha = 0.3$.}} 
    \label{fig: effective potential plots for different Rs}
\end{figure}

It is also worth analyzing the effect of the quantum core size on the effective potential. The parameter $R_{\rm s}$ is associated with the black hole inner region, and as already mentioned, the event horizon depends on its value, as depicted in Fig. \ref{fig:horizon_radius}. Recalling that now we consider $R_{\rm M} = 2$, one can realize that the major changes in the horizon position occur for quantum cores of order $R_{\rm s} \sim 0.8$ or higher. Fig.  \ref{fig: effective potential plots for different Rs} depicts the effect of higher $R_{\rm s}$ values on the effective potential curves. In both panels, we fixed $\alpha = 0.3$. {We can clearly see that for progressively larger quantum core sizes, the effective potential barrier becomes increasingly asymmetric. \textcolor{black}{This behavior reflects the fact that the coherent quantum corrections mainly modify the near-horizon region of the geometry, while the asymptotic Schwarzschild-like behavior remains essentially unchanged. Consequently, only the inner side of the potential barrier undergoes significant deformation.} In particular, the left-hand side of the barrier broadens, producing a more gradual rise toward the maximum, while the right-hand side remains almost unchanged. As a consequence, the location of the peak is slightly shifted toward larger values of the tortoise coordinate, although its maximum value is only marginally affected. These modifications become appreciable only when $R_{\rm s}$ approaches the regime where the event horizon begins to depart significantly from the Schwarzschild value, consistently with the behavior shown in Fig. \ref{fig:horizon_radius}. Therefore, the dominant influence of the quantum core size does not alter the height of the potential barrier, but rather modifies its shape and width, especially in the near-horizon region. Since the QNM spectrum is determined by the shape of the effective potential, these geometric modifications are expected to induce quantitative changes in the oscillation frequencies and damping times of the coherent quantum hairy black hole. Such changes are expected to affect the propagation of perturbations through the geometry and may lead to quantitative differences in the corresponding QNM spectrum.}

\blt{Comparing the two multipoles in Fig. \ref{fig: effective potential plots for different Rs}, we observe that the influence of the quantum core size is qualitatively the same for both $l=1$ and $l=2$. In both cases, increasing $R_{\rm s}$ yields the effective potential barrier to become progressively more asymmetric, with the left-hand side broadening and the peak shifting slightly toward larger values of the tortoise coordinate, while the right-hand side remains essentially unchanged. The maximum of the potential is only weakly affected, indicating that the dominant effect of the quantum core is to reshape the barrier rather than to modify its height. Quantitatively, however, the deformation is slightly more pronounced for the $l=2$ mode, whose higher potential barrier exhibits a somewhat larger broadening of the left shoulder and a marginally greater displacement of the peak. Nevertheless, the similarity between the two multipoles suggests that the quantum core primarily modifies the near-horizon geometry, with only a weak dependence on the angular momentum number. Consequently, one expects the corresponding QNM spectra to exhibit the same qualitative trends for different multipoles, although the precise frequency shifts and damping rates must be determined through the QNM analysis presented in the next section.}

\subsection{The WKB approximation}
\textcolor{black}{Since the effective potential consists of a single smooth barrier, the WKB approximation provides an efficient semi-analytical method for computing the quasinormal frequencies.}
To find the QNM frequencies of the coherent quantum GD hairy black hole, we employ the WKB approximation. The quasinormal frequencies are given by
\begin{eqnarray}
    \omega^2 = \left[V + \sqrt{(-2V^{(2)}} \Gamma\right] - i \sqrt{\delta (-2V^{(2)})} \,(1 + \Omega)\, ,
\end{eqnarray}
where
\begin{eqnarray}
  \!\!\!  \!\!\!\!\!\!\!\!\!\Gamma &=& \frac{1}{\sqrt{-2 V^{(2)}}} \left[\frac{1}{8} \left(\frac{V^{(4)}}{V^{(2)}}\right) \left(\frac{1}{4} + \delta \right)
    - \frac{1}{288}\left(\frac{V^{(3)}}{V^{(2)}}\right)^2 (7 + 60 \delta)   \right]\, ,\nonumber\\
  \!\!\!\!\!\!\!\!\!\!\!\!  \Omega &=& -\frac{1}{2V^{(2)}} \Biggl[\frac{5}{6912} \left(\frac{V^{(3)}}{V^{(2)}}\right)^4 (77 + 188\delta) - \frac{1}{134}\left(\frac{(V^{(3)})^2 V^{(4)}}{(V^{(2)})^3} (51 + 100\delta)\right)\nonumber \\
   \!\!\!\!\!\!\!\!\!\!\!\! &&+ \frac{1}{2304} \left(\frac{V^{(4)}}{V^{(2)}}\right)^2 (67 + 68\delta)
    +\frac{1}{288} \left(\frac{V^{(3)} V^{(5)}}{(V^{(2)})^2}\right)(19 + 28\delta) - \frac{1}{288} \left(\frac{V^{(6)}}{V^{(2)}}\right) (5 + 4 \delta)  \Biggl]\, 
\end{eqnarray}
and
\begin{eqnarray}
    \delta = \left(\frac{1}{2} + n\right)^2\, .
\end{eqnarray}
The terms $V^{(k)}$ denote the $k^{\rm th}$ derivative of the effective potential,
\begin{eqnarray}
    V^{(k)} = \frac{d^k V}{dr_\star^k}\, ,
\end{eqnarray}
evaluated at the maximum of the potential. We present results for the QNMs frequencies for values of $l$ such that $n \leq l$, since this is the range where the WKB method is most reliable. \textcolor{black}{For lower multipoles, particularly the fundamental $l=0$ mode, the WKB approximation is generally less accurate. Nevertheless, it remains sufficient for identifying the qualitative dependence of the spectrum on the parameters $\alpha$ and ${R}_{\rm s}$.} Different cases are analyzed, aiming to illustrate the influence of $\alpha$ and $R_{\rm s}$ on the oscillation frequencies, and how they deviate from classical black hole models, including the classical hairy GD black hole.

The first case analyzed is the coherent quantum hairless black hole, consisting of setting $\alpha = 0$ in Eq. \eqref{eq: metric function for qnm}, which yields a metric function that reads
\begin{eqnarray}
    f(r,R_{\rm s}) = 1 -\frac{2M}{r} \,\operatorname{erf}\left(\frac{r}{R_{\rm s}}\right)\, .
\end{eqnarray}
Such a metric function corresponds to a quantum Schwarzschild geometry obtained by the coherent states approach. The QNMs of such a configuration are obtained in Ref. \cite{Antonelli:2026qnm}, where the authors implement the WKB method using Padé approximants of order [6/7]. Their calculation is based on the library routine presented in Ref. \cite{Konoplya:2019hlu}. We employ the third-order WKB approximation to calculate the quasinormal frequencies for different values of the quantum core size $R_{\rm s}$. These results can be seen in Table \ref{tab:qnm alpha = 0}. The column $R_{\rm s} = 0$ corresponds to the classical Schwarzschild black hole. We notice that for a quantum core size of $R_{\rm s} = 0.5$ there is no relevant deviation for the quasinormal frequencies compared to the classical hairy black hole configuration. This is in agreement with our previous remarks, considering the effects of low $R_{\rm s}$ values on the effective potential. 

\begin{table}[h!]
    \centering
    \setlength{\tabcolsep}{12pt}
    \caption{QNM frequencies ($2M\omega$) for the hairless ($\alpha = 0$) coherent quantum Schwarzschild black hole.}
    \label{tab:qnm alpha = 0}
    \begin{tabular}{llllll}
        \specialrule{0.15em}{0pt}{2pt}
        $l$ &
        $n$ &
        $R_{\rm s} = 0$ (Classical) &
        $R_{\rm s} = 0.5$  &
        $R_{\rm s} = 1.0$  &
        $R_{\rm s} = 1.5$  \\
        \specialrule{0.08em}{2pt}{2pt}

        0 & 0 & $0.2093 - 0.2304i$ & $0.2093 - 0.2304i$ & $0.1758-0.2154i$ & $0.1825-0.2139i$  \\
        1 & 0 & $0.5822 - 0.1960i$ & $0.5822 - 0.1960i$ & $0.5801-0.1945i$ & $0.5782-0.1762i$ \\
        1 & 1 & $0.5244 - 0.6149i$ & $0.5244 - 0.6149i$ & $0.5109-0.6106i$ & $0.4949-0.5634i$\\
        2 & 0 & $0.9664 - 0.1936i$ & $0.9664 - 0.1936i$ & $0.9656-0.1930i$ & $0.9694-0.1772i$\\
        2 & 1 & $0.9264 - 0.5916i$ & $0.9264 - 0.5916i$ & $0.9216-0.5892i$ & $0.9222-0.5412i$\\
        2 & 2 & $0.8633 - 1.007i$  & $0.8633 - 1.007i$  & $0.8493-1.0020i$ & $0.8422-0.9238i$\\
        3 & 0 & $1.350 - 0.1930i$  & $1.350 - 0.1930i$  & $1.3500-0.1926i$ & $1.3570-0.1775i$\\
        3 & 1 & $1.321 - 0.5847i$  & $1.321 - 0.5847i$  & $1.3180-0.5828i$ & $1.3250-0.5369i$\\
        3 & 2 & $1.270 - 0.9882i$  & $1.270 - 0.9882i$  & $1.2610-0.9842i$ & $1.2670-0.9067i$\\
        3 & 3 & $1.204 - 1.402i$   & $1.204 - 1.402i$   & $1.1860-1.3960i$ & $1.1900-1.2870i$\\
        \specialrule{0.15em}{2pt}{0pt}
    \end{tabular}
\end{table}

For a non-vanishing value of  $\alpha$, we recover the metric \eqref{eq: metric function for qnm}. Computing the WKB approximation to calculate the quasinormal frequencies for $\alpha =0.3$, we obtain the results shown in Table \ref{tab:qnm alpha = 0.3} for different values of $R_{\rm s}$. Table \ref{tab:qnm alpha = 0.6} shows the results for the quasinormal frequencies when considering $\alpha = 0.6$. The columns $R_{\rm s} = 0$ correspond to the classical GD configuration, where there are no quantum corrections to the black hole geometry. The first thing to notice from the results shown in Tables \ref{tab:qnm alpha = 0.3} and \ref{tab:qnm alpha = 0.6} is that a non-vanishing $\alpha$ results in relevant changes to the quasinormal frequencies when compared to the hairless configuration (Table \ref{tab:qnm alpha = 0}), even for lower values of $R_{\rm s}$. We also verify that for higher values of $\alpha$, both the real and imaginary frequencies decrease. Since the imaginary frequency oscillation is associated with the damping rate, this implies that the GD hair parameter $\alpha$ brings the black hole to a behavior closer to an oscillator.    \textcolor{black}{Since the magnitude of the imaginary part decreases, the damping time $\tau=1/|{\rm Im}\,\omega|$ increases. Therefore, perturbations survive for longer times, indicating that the GD hair makes the black hole ring down more slowly.}

\begin{table}[h!]
    \centering
    \setlength{\tabcolsep}{12pt}
    \caption{QNM frequencies ($2M\omega$) for the coherent quantum GD hairy black hole with $\alpha = 0.3$.}
    \label{tab:qnm alpha = 0.3}
    \begin{tabular}{llllll}
        \specialrule{0.15em}{0pt}{2pt}
        $l$ &
        $n$ &
        $R_{\rm s} = 0$ (GD) &
        $R_{\rm s} = 0.5$  &
        $R_{\rm s} = 1.0$  &
        $R_{\rm s} = 1.5$  \\
        \specialrule{0.08em}{2pt}{2pt}

        0 & 0 & $0.2028-0.2143i$ & $0.1939-0.2169i$ & $0.1688-0.2062i$ & $0.1671-0.2027i$ \\
        1 & 0 & $0.5727-0.1869i$ & $0.5475-0.1843i$ & $0.5455-0.1828i$ & $0.5419-0.1671i$ \\
        1 & 1 & $0.5199-0.5828i$ & $0.4907-0.5787i$ & $0.4793-0.5751i$ & $0.4568-0.5358i$ \\
        2 & 0 & $0.9491-0.1849i$ & $0.9097-0.1822i$ & $0.9088-0.1815i$ & $0.9104-0.1689i$ \\
        2 & 1 & $0.9132-0.5637i$ & $0.8710-0.5569i$ & $0.8667-0.5543i$ & $0.8630-0.5158i$ \\
        2 & 2 & $0.8554-0.9574i$ & $0.8096-0.9479i$ & $0.7977-0.9436i$ & $0.7814-0.8807i$ \\
        3 & 0 & $1.3260-0.1844i$ & $1.2710-0.1817i$ & $1.2710-0.1812i$ & $1.2750-0.1694i$ \\
        3 & 1 & $1.2990-0.5579i$ & $1.2430-0.5505i$ & $1.2400-0.5484i$ & $1.2430-0.5124i$ \\
        3 & 2 & $1.2530-0.9416i$ & $1.1930-0.9304i$ & $1.1860-0.9263i$ & $1.1850-0.8651i$ \\
        3 & 3 & $1.1930-1.3340i$ & $1.1300-1.3200i$ & $1.1150-1.3140i$ & $1.1050-1.2280i$ \\
        \specialrule{0.15em}{2pt}{0pt}
    \end{tabular}
\end{table}


\begin{table}[h!]
    \centering
    \setlength{\tabcolsep}{12pt}
    \caption{QNM frequencies ($2M\omega$) for the coherent quantum GD hairy black hole with $\alpha = 0.6$.}
    \label{tab:qnm alpha = 0.6}
    \begin{tabular}{llllll}
        \specialrule{0.15em}{0pt}{2pt}
        $l$ &
        $n$ &
        $R_{\rm s} = 0$ (GD) &
        $R_{\rm s} = 0.5$  &
        $R_{\rm s} = 1.0$  &
        $R_{\rm s} = 1.5$  \\
        \specialrule{0.08em}{2pt}{2pt}

        0 & 0 & $0.1969-0.1978i$ & $0.1803-0.2043i$ & $0.1597-0.1960i$ & $0.1523-0.1923i$ \\
        1 & 0 & $0.5630-0.1779i$ & $0.5165-0.1736i$ & $0.5147-0.1725i$ & $0.5095-0.1585i$ \\
        1 & 1 & $0.5160-0.5512i$ & $0.4610-0.5455i$ & $0.4518-0.5432i$ & $0.4221-0.5098i$ \\
        2 & 0 & $0.9315-0.1762i$ & $0.8588-0.1719i$ & $0.8579-0.1713i$ & $0.8578-0.1611i$ \\
        2 & 1 & $0.8999-0.5361i$ & $0.8215-0.5251i$ & $0.8178-0.5233i$ & $0.8100-0.4919i$ \\
        2 & 2 & $0.8478-0.9083i$ & $0.7621-0.8939i$ & $0.7521-0.8910i$ & $0.7267-0.8398i$ \\
        3 & 0 & $1.3010-0.1758i$ & $1.2000-0.1714i$ & $1.2000-0.1710i$ & $1.2010-0.1618i$ \\
        3 & 1 & $1.2770-0.5313i$ & $1.1730-0.5192i$ & $1.1710-0.5177i$ & $1.1700-0.4893i$ \\
        3 & 2 & $1.2360-0.8954i$ & $1.1260-0.8775i$ & $1.1190-0.8746i$ & $1.1110-0.8258i$ \\
        3 & 3 & $1.1820-1.2670i$ & $1.0640-1.2450i$ & $1.0520-1.2410i$ & $1.0300-1.1720i$ \\
        \specialrule{0.15em}{2pt}{0pt}
    \end{tabular}
\end{table}

\textcolor{black}{Our analysis shows that the GD hair and the coherent quantum core leave complementary imprints on the QNM spectrum. While the hair parameter primarily lowers the effective potential barrier, reducing both the oscillation frequency and the damping rate, the quantum core mainly reshapes the near-horizon portion of the potential and becomes relevant only when its size approaches the horizon scale. Consequently, the two parameters modify the ringdown signal in qualitatively different ways. The GD hair mainly shifts the characteristic frequencies of the oscillations, whereas the coherent quantum core produces additional changes associated with the deformation of the scattering potential close to the horizon. Since the ringdown phase of a compact-object merger is entirely governed by the QNM spectrum, sufficiently precise GW observations could, in principle, discriminate between classical scalar hair and coherent quantum corrections. Although the present analysis is restricted to scalar perturbations, the observed trends originate from modifications of the background geometry itself and are therefore expected to carry over, at least qualitatively, to the physically relevant gravitational perturbations.}

\section{Conclusions}
\label{sec6}
Applying this prescription to the GD hairy black hole yields a geometry that is asymptotically classical while developing a finite quantum core at short distances. The Gaussian regulator suppresses high-momentum modes in the Fourier--Bessel decomposition, eliminating the characteristic $r^{-1}$ and $r^{-2}$ singularities. As a result, the coherent quantum GD hairy metric function becomes analytic at the origin, with all curvature invariants remaining finite under suitable parameter conditions. 

The classical GD hairy potential decomposes into an RN term and a Yukawa-like contribution controlled by the hair parameter $\alpha$ and the scale $\ell$. After Gaussian regularization, both the electromagnetic and hairy sectors are finite at the origin. The resulting coherent quantum GD hairy spacetime smoothly interpolates between an asymptotically flat exterior and a nonsingular quantum core, preserving the large-distance behavior while resolving the central singularity. 
From the effective Einstein equations, the corresponding stress-energy tensor can be interpreted as an anisotropic quantum fluid. The energy density and pressures remain finite at the center, with a vacuum-like equation of state analogous to a de Sitter core. The difference between tangential and radial pressures reflects the intrinsic anisotropy induced by quantum smearing and hair contributions.

We also computed the Misner--Sharp mass function and showed that, while the quantum RN sector asymptotically reproduces the ADM mass $M$, the hairy corrections renormalize the total mass at infinity. For small values of $R_{\rm s}$, the mass function reduces to the classical GD hairy expression with additional Yukawa-type corrections. 
The horizon structure of the coherent quantum GD hairy solution was investigated numerically. Depending on the values of $(R_{\rm M}, R_{\rm Q}, R_{\rm s}, \alpha)$, the coherent quantum GD hairy spacetime can exhibit two horizons, a degenerate (extremal) configuration, or no horizon at all. The competition between charge, mass, quantum core size, and hair amplitude determines whether a black hole or a horizonless compact object forms. In particular, we derived the condition for regularity at the origin and identified parameter regions where the geometry is nonsingular.

We analyzed null geodesics in the quantum-corrected GD hairy background. The photon ring radius, critical impact parameter, and light deflection were computed numerically. Both the quantum core scale $R_{\rm s}$ and the hair parameter $\alpha$ produce measurable deviations from the classical and quantum RN cases,  in the strong-field regime. These modifications leave imprints on black hole shadows and gravitational lensing, providing potential phenomenological settings to constrain the scale of quantum corrections and the presence of hair. {We additionally employed the third-order WKB method to calculate the QNMs of the coherent quantum GD hairy black hole for scalar perturbations, under the saturated hair charges conditions (Eq. \eqref{eq: lower bounds over l}), reducing its metric function to \eqref{eq: quantum + hairy metric}, which leads to a Schwarzschild-like coherent quantum GD hairy black hole. The results displayed in Tables \ref{tab:qnm alpha = 0}, \ref{tab:qnm alpha = 0.3}, and \ref{tab:qnm alpha = 0.6} show that both parameters $R_{\rm s}$ and $\alpha$ modify the oscillation frequencies relative to both the classical Schwarzschild black hole and the GD hairy black hole. These findings indicate that the quantum corrections introduced through the coherent-state approach also leave observable imprints on the dynamics of scalar perturbations in this spacetime. Since the third-order WKB approximation provides only a first estimate of the QNM spectrum, more accurate frequencies are expected from higher-order WKB schemes or fully numerical methods. We further expect that gravitational and other higher-order spin perturbations exhibit analogous deviations from their classical counterparts, potentially providing an additional observational probe of the model in the era of next-generation GW  detectors.}

The coherent quantum GD hairy black hole setup provides a self-consistent framework in which UV regularization, singularity resolution, modified horizon structure, and potentially observable strong-field signatures arise from a quantum description. The Gaussian smearing scale $R_{\rm s}$ governs the transition between classical geometry and a regular quantum core, while the hair parameter $\alpha$ controls deviations from the purely quantum-corrected RN case. 
Our results support the interpretation of GD hairy black holes as macroscopic quantum states whose classical singularities are softened, while admitting phenomenologically relevant hair.

\medbreak
{\textbf{Acknowledgements}}: 
HN is financed by the Coordination for the Improvement of Higher Education Personnel - Brazil (CAPES) - Finance Code 001. RdR is supported by The S\~ao Paulo Research Foundation (FAPESP) 
(Grants No. 2025/23004-9, No. 2021/01089-1, and No. 2024/05676-7) and the National Council for Scientific and Technological Development (CNPq) (Grants No. 303742/2023-2 and No. 401567/2023-0).

\renewcommand{\bibfont}{\footnotesize}
\bibliographystyle{apsrev}
\bibliography{gdq}

\end{document}